\documentclass{aastex63}
\usepackage{float}
\usepackage{graphicx}  
\graphicspath{{./}} 

\newcommand{\hone}{H\,{\footnotesize{I}}}  	
\newcommand{\hii}{H\,{\footnotesize{II}}}	
\newcommand{\ppa}{Pegasus-Pisces Arch}		
\newcommand{\los}{line-of-sight}		

\shorttitle{The Long Tails of the Pegasus-Pisces Arch}
\shortauthors{Shelton et al.}

\begin{document}

\title{The Long Tails of the Pegasus-Pisces Arch Intermediate Velocity Cloud}

\author{R. L. Shelton}
\email{rls@physast.uga.edu}
\author{M. E. Williams}
\email{elliott.williams14@gmail.com}
 \author{M. C. Parker}
  \email{mcparker225@gmail.com}
\author{J. E. Galyardt}
\email{jason.galyardt@gmail.com}
\author{Y. Fukui}
\email{fukui@a.phys.nagoya-u.ac.jp}
\author{K. Tachihara}
\email{k.tachihara@a.phys.nagoya-u.ac.jp}

\date{\today}



\begin{abstract}

  We present hydrodynamic simulations of the \ppa\ (PP Arch), an intermediate velocity cloud in our Galaxy.
The PP Arch, also known as IVC 86-36, 
is unique among intermediate and high velocity clouds, because its twin tails are unusually long and narrow.
Its $-50$~km~s$^{-1}$ \los\ velocity qualifies it as an intermediate velocity cloud, but the tails' orientations
  indicate that the cloud's total three-dimensional speed is at least $\sim100$~km~s$^{-1}$.
  This speed is supersonic in the Reynold's Layer and thick disk.
  We simulated the cloud as it travels supersonically through the Galactic thick and thin disks
  at an oblique angle relative to the midplane.
  Our simulated clouds grow long double tails and
  reasonably reproduce the \hone\ 21~cm intensity and velocity of the head of the PP Arch.  
  A bow shock protects each simulated cloud from excessive shear
  and
  lowers its Reynolds number.
  These factors
  may similarly protect the PP Arch and enable the survival of its unusually long tails.
  The simulations predict the future hydrodynamic behavior of the cloud when it collides with denser gas nearer to the
  Galactic midplane.   It appears that the PP Arch's fate is to deform, dissipate, and merge with the Galactic disk.
  \\

%

  \noindent
{\it{Unified Astronomy Thesaurus concepts:}} \ Interstellar clouds (834); High-velocity clouds (735); Computational astronomy (293); Hydrodynamical simulations (767)
  
\end{abstract}




\section{Introduction}
\label{sect:intro}

\begin{figure*}[htb!]		 
\epsscale{1}

{(a)}\includegraphics[scale=0.8]{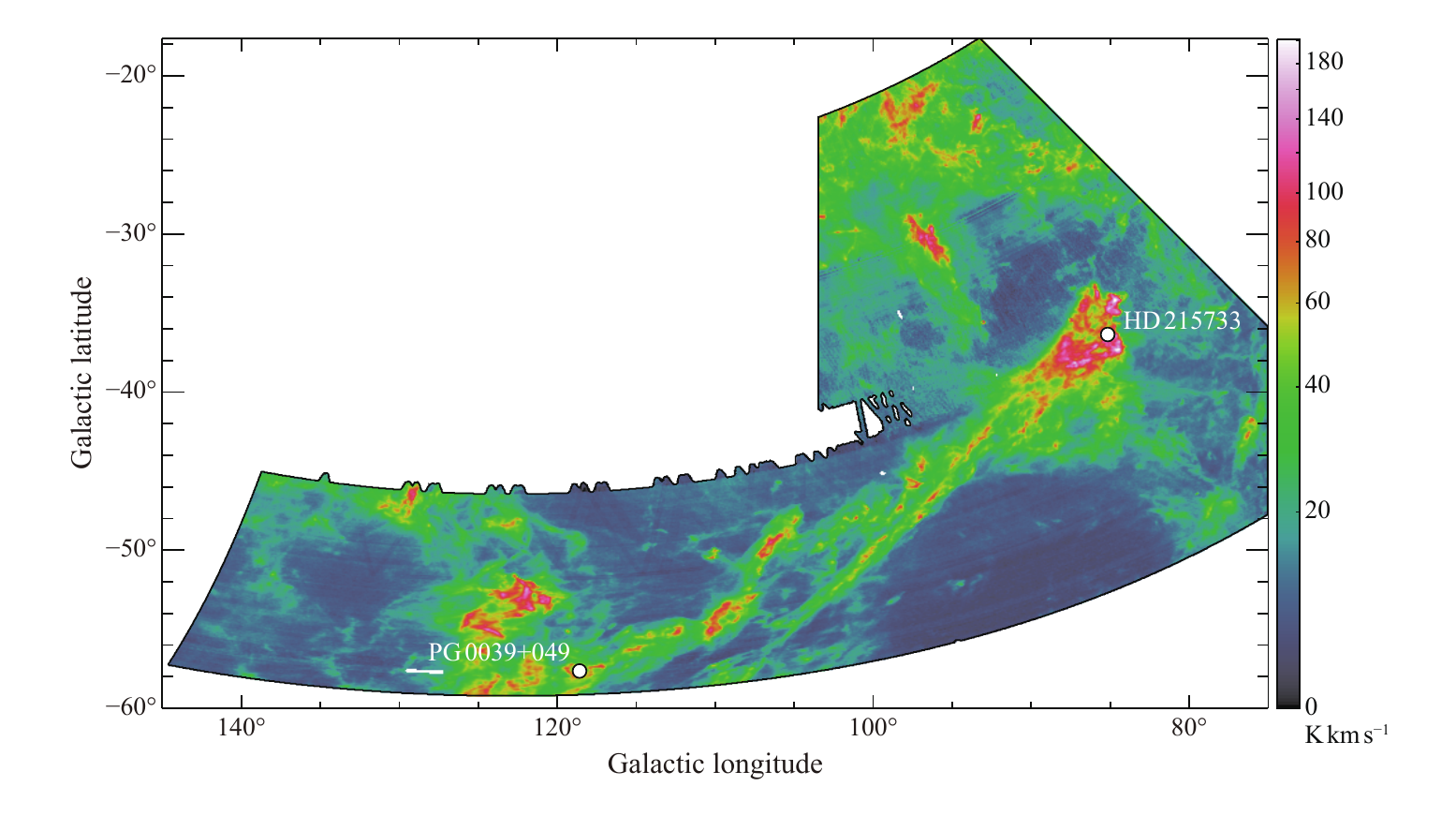} \\
{(b)}\includegraphics[trim=20 200 200 200, scale=0.4]{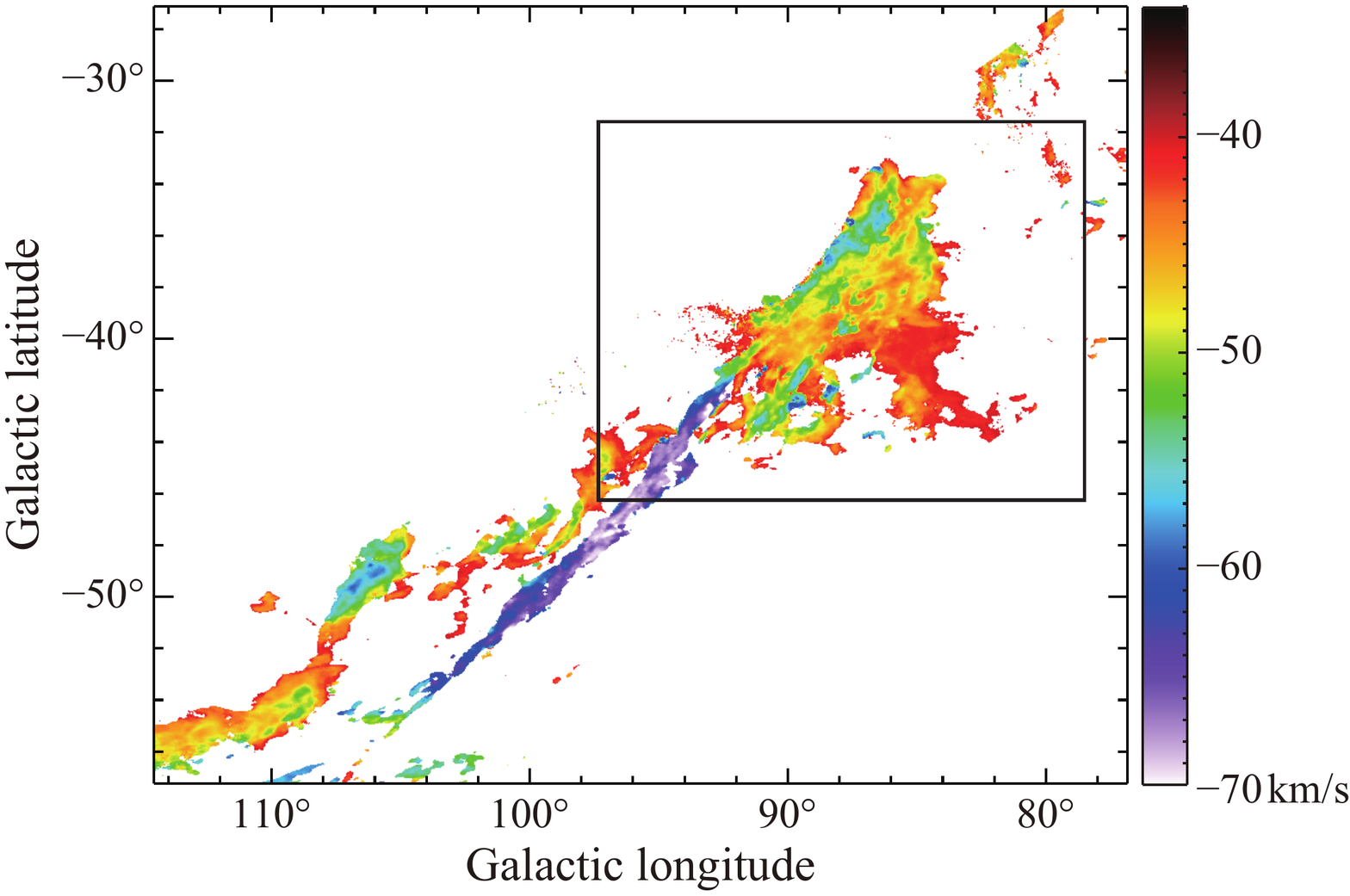}  \hspace*{3cm} 
{(c)}\includegraphics[scale=1.05]{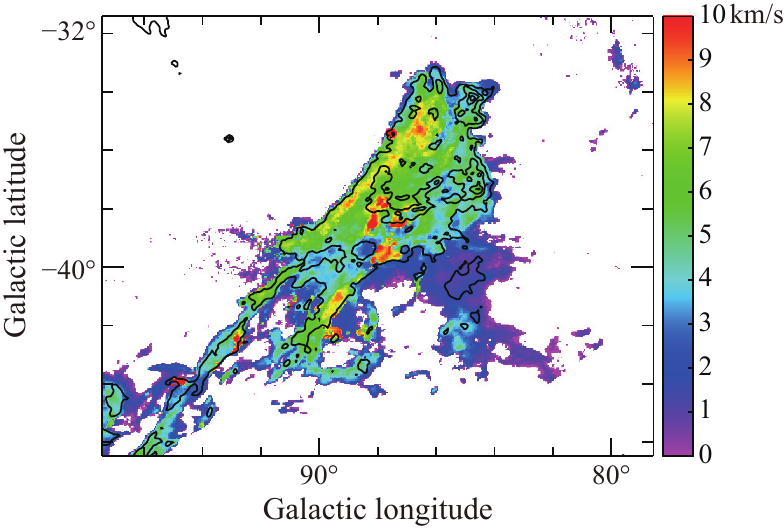}

\caption{ (a) Map of \hone\ 21~cm emission intensity integrated over the velocity range of
  $-80$~km~s$^{-1}$ to $-30$~km~s$^{-1}$ for the \ppa\ region.
  The southeastern ends of the tails continue somewhat beyond the extent of this map.
  The stars HD 215733, at $\ell = 85.2^{\rm{o}}$, $b = -36.4^{\rm{o}}$,
  and PG 0039+049, at $\ell = 118.59^{\rm{o}}$, $b = -57.64^{\rm{o}}$, are marked on the map.
  Their spectra show signs of absorption by the \ppa.
  Figure reproduced from Fukui et al. (2021). 
%
  (b) \hone\ first moment map showing the LSR velocities   
    between $-35$~km~s$^{-1}$ and $-70$~km~s$^{-1}$.
    Figure adapted from Fukui et al. (2021) by the addition of a smaller box which outlines the
    region examined in the subsequent map.
    (c) Second moment map showing the velocity dispersion of the head of the \ppa\
    within the smaller region outlined in the first moment map.
    The overlaid contours show the distribution of
    velocity-integrated intensity (i.e., the zeroth moment).  The contour levels are incremented by 50~K~km~s$^{-1}$.
    This map was made from GALFA-\hone\ Data Release 1 data (Peek et al. 2011).
}
\label{fig:fukui_hi_map}
\end{figure*}

\hone\ maps of the Pegasus-Pisces region reveal an unusually long, narrow, and straight cloud of
intermediate velocity
gas extending $\sim42^{\rm{o}}$,
from $(l,b) \sim (84^{\rm{o}},-34^{\rm{o}})$ to $(l,b) \sim (130^{\rm{o}},-62^{\rm{o}})$,
with an apparent width of $\sim3^{\rm{o}}$.    It has a distinct head and tapered, bifurcated tail.
It appears clearly in the map of \hone\ with velocities between $V_{\rm{LSR}} = -85$  and $-45$~km~s$^{-1}$ 
in Figure 17 of Wakker (2001)
and the maps of \hone\ with velocities between $V_{\rm{LSR}} \sim -80$ and $\sim -30$~km~s$^{-1}$ in Figures 2 - 4 in
Fukui et al. (2021).   Their Figure~2 is reproduced here as our Figure~\ref{fig:fukui_hi_map}.
Wakker (2001) identified this object as an intermediate velocity cloud
(IVC).   Wakker (2001) defined IVCs as clouds with local standard of rest velocities
    of $V_{\rm{LSR}} = \sim40$~km~s$^{-1}$ to 90~km~s$^{-1}$, but the upper $V_{\rm{LSR}}$ cut-off has been
    placed as high as 100~km~s$^{-1}$ by some authors (e.g., Richter 2017).
    Wakker (2001) named this especially long IVC after its location,
calling it
%
the Pegasus-Pisces Arch, 
abbreviated as the
PP Arch.
Fukui et al. (2021), who concentrated
on the portion that runs from $(l,b) \sim (84^{\rm{o}},-34^{\rm{o}})$ to $(l,b) \sim (110^{\rm{o}},-55^{\rm{o}})$,
named it after the Galactic coordinates of its head, hence IVC 86-36.   
%

%
%
%

The \ppa\ is long, but pennant-shaped,
with a broader head and narrower, tapered tails.
Its relatively streamline shape
contrasts clearly with those of its irregularly shaped low velocity neighbors
MBM 53, HLCG 92-35, MBM 54, and MBM 55
(Magnani et al. 1985; Yamamoto et al. 2003; Fukui et al. 2021),
which, together, resemble a curved archipelago of small and midsized islands 
in maps of \hone, CO, and tracers of dust.

The \ppa\ is also more streamlined than other intermediate velocity clouds.
The best known IVCs, namely the
IV Arch, Low Latitude IV Arch, and IV Spur, are ovular and notably clumpy
(Danly 1989, Kuntz $\&$ Danly 1996, Wakker 2001, Richter 2017).
%
Complex GP is a circular collection of clumps (Wakker 2001; Richter 2017).
Complex K and Complex L
are wide swaths of clumps (see Figure 16 of Wakker 2001;
Haffner et al. 2001; Richter 2017).
Complex L includes both intermediate velocity gas and
high velocity gas, which is defined as that with $V_{\rm{LSR}}$ greater
than 90 or 100 km~s$^{-1}$ (Wakker 2001; Richter 2017).
%
%

Likewise, Complexes C and M, the Leading Arm, and the other high velocity cloud (HVC) complexes
appear broader, clumpier, or less organized than the Pegasus-Pisces Arch
(Wakker \& van Woerden 1997; Westmeier 2018;
see also Kalberla et al. 2005).
The Magellanic Stream is an exceptional case, as it is shaped like entwined ribbons of
gas left behind by the SMC and LMC
(Richter et al. 2013; Fox et al. 2013; Fox et al. 2014; see also Nidever et al. 2008). 
Some individual clouds within the Magellanic Stream have been described as {\it{head-tail}}
in which one end of the cloud is broader and has greater column density than the other end
(Br\"{u}ns et al. 2000).
Some head-tail clouds also exhibit velocity gradients (For et al. 2014).
However, these clouds are not very long or narrow.
About a third of the compact and semi-compact HVCs studied by
Putman et al. (2011) have been described as head-tail clouds,
but these clouds are far more ovular than the Pegasus-Pisces Arch.
An interesting example of a head-tail compact HVC is HVC125+41-207, which 
has an aspect ratio of 3 or 4.
It has a teardrop shape in low resolution \hone\ maps,
but 
high resolution observations reveal that its head is actually
composed of three highly irregular \hone\ clumps
(Br\"{u}ns et al. 2001).
%

Aside from its unusual tail, the \ppa\ shares some similarities with other IVCs.
Like most IVCs (R\"{o}hser et al. 2016), the \ppa\ is located within 2~kpc
of the Galactic midplane, but is not in the midplane.
Examples of such clouds include
%
the largest intermediate velocity complex, the IV Arch, which is located between 800 and 1500~pc above the midplane (Kuntz \& Danly 1996),
and the IV Spur, which is located between 1200 and 2100~pc above the midplane (Kuntz \& Danly 1996).

Like the majority of IVCs (Richter 2017), the \ppa\ has a negative \los\ velocity and is probably
falling toward the Galactic disk.
Its \los\ velocity is approximately $-50$~km~s$^{-1}$
and
the cloud is oriented with its head nearer to the Galactic midplane than is its tail.
If the cloud is moving in the direction of its long axis, as assumed (Fukui et al. 2021), then
it is traveling toward the Galactic disk 
at a $\sim45^{\rm{o}}$ angle with respect to the Galactic midplane and is probably traveling with a total
velocity of $\sim 100$~km~s$^{-1}$.

The \ppa\ probably originated well beyond the Galactic disk.
Absorption spectroscopy of the cloud's head yields a metallicity of
0.54 $\pm$ 0.04 solar (Wakker et al., 2001),
while emission spectroscopy of Planck and IRAS data of the head yields an upper limit of $\sim0.2$ solar (Fukui et al. 2021,
also see Fukui et al. 2015 for a discussion of the relationship between 353~GHz emission and dust content in clouds).
Having a substantially subsolar metallicity is neither unique nor ubiquitous among IVCs
(Wakker 2001, Hernandez et al. 2013),
but is uncommon and does
suggest that the cloud originated outside of our Galaxy and has fallen into it (Fukui et al. 2021).
Contemplations of the cloud's past naturally lead to contemplations of its future.
Will the cloud come to rest in the Galactic disk, punch through 
like larger, faster simulated clouds (Tepper-Garc\'{i}a \& Bland-Hawthorn 2018) or dissipate (Galyardt \& Shelton 2016)?

Computer simulations can shed light on the situation.
Already, an array of simulations have been performed for
IVCs' high speed cousins, HVCs.
Model HVCs moving through the gradiated density gas within a few kpc of the Galactic midplane develop
smooth tails.
The tails are short and stocky in most simulations
(see Santillan et al. 1999;
Santillan et al. 2004;
Jel\'{i}nek \& Hensler 2011),
but are longer when the fairly massive clouds are simulated (Galyardt \& Shelton 2016).
Tails also grow on simulated HVCs traveling through very hot, low density gas, like that expected much
farther from the Galactic midplane (see Heitsch \& Putman 2009;
Kwak et al. 2011;
Gritton et al. 2014;
Armillotta et al. 2017;
Gritton et al. 2017; Sander \& Hensler 2020),
but 
these tails are generally much more globular and erratic than those on simulated HVCs nearer to the midplane and much blobbier than
the \ppa.
Far less simulational work has been done on IVCs.
An exception is
Kwak et al. (2009),
who modeled clouds that accelerate from
zero velocity as they fall through the gradiated density gas above the Galactic midplane.
Short tails develop on some of their simulated clouds.
The following simulations will help to broaden the understanding of infalling clouds and shed light on the development
of the long tails of the \ppa.


      We performed
      a suite of
      bespoke simulations of the \ppa\ cloud.
      We used the observed \los\ velocity of the head, the orientation of the cloud's head-tail structure, and the head's
\hone\ column density to guide our choice of input parameters for the simulations and to select good models from the
set of preliminary simulations.
The observations and the resulting constraints on the models 
are listed in Section~\ref{sect:observations}.
We model the IVC's hydrodynamic interactions with its surrounding Galactic 
environment, using the FLASH simulation framework (Fryxell, 2000).
The FLASH code, domain geometry, and input parameters are discussed in 
Section~\ref{sect:simulations}.
The simulated clouds develop long, smooth, twinned tails at approximately the observed inclination angle
and agree with the head's observed \hone\ intensity and velocity.
This is shown in
Section~\ref{sect:results} where we present the simulational models and compare them with the observations.
The clouds instigate bow shocks that reduce the shear between the tails and the surrounding gas.
The Reynolds number is low, which portends little turbulence and allows the tails to grow relatively undisturbed.
In Section~\ref{sect:discussion}, we discuss this issue and the viewing geometry as
possible reasons why
the tails of the \ppa\ are long and relatively smooth while the tails of most IVCs and HVCs are not.
We summarize the key points in Section~\ref{sect:summary}.

\section{Observed Characteristics}
\label{sect:observations}

In order to construct simulational models of the cloud,
we need to consider the \ppa's observed size, \hone\ mass, orientation, velocity, and height above the Galactic midplane, $z$.
The latter is used in Section~\ref{sect:simulations}
to estimate the gravitational acceleration and density of ambient material in the vicinity
of the cloud.    To this list, we add the distance
to the cloud, as it factors into the cloud's size, mass, and distance from the midplane.
We also add the \hone\ intensity of the cloud's head, the overall shape of the cloud, and the velocity dispersion of the cloud's head,
as they have been observed (Fukui 2021; Wakker 2001)
and can be used to test the simulational models.

We begin with the distance to the cloud, as it factors into so many other quantities.
The distance to the cloud is constrained by two stars.
The star HD 215733, at 
$\ell = 85.2^{\rm{o}}$, $b = -36.4^{\rm{o}}$, is within the head's footprint (see Figure~\ref{fig:fukui_hi_map}).
Its spectrum includes 
several absorption lines of low ionization species within the velocity range of the cloud
(Fitzpatrick \& Spitzer 1997),
and therefore, 
the upper limit on the distance to the cloud's head is equal
to the distance to the star.
A spectroscopic analysis finds the star's distance to be $\sim2900$~pc (Fitzpatrick \& Spitzer 1997),
   while a parallax analysis of GAIA data finds it to be $3.5 \pm 0.9$ kpc (Fukui et al. 2021).
%
   As also shown in Figure~\ref{fig:fukui_hi_map}, the star PG 0039+049, at $\ell = 118.59^{\rm{o}}$, $b = -57.64^{\rm{o}}$, is within
   the footprint of one of the tails.   
%
Centurion et al. (1994) discovered intermediate velocity absorption features in the star's spectrum and
Smoker et al. (2011) 
confirmed that the star places a firm upper limit on the cloud's distance.
Moehler et al. (1990)
determined the star's distance to be $1050 \pm 400$~pc.

Upper limits on the $|z|$ of the cloud and its projected distance in the Galactic plane can be
easily calculated from the stellar distances.
The distance to star HD 215733 multiplied by the star's $\cos(b)$ yield
upper limits on the cloud's projected distance in the Galactic midplane of 2330~pc and $2820 \pm 720$~pc.
A similar calculation using the distance to PG 0039+049 yields
a much smaller upper limit on the cloud's projected distance in the Galactic midplane:  $560 \pm 210$~pc.
The $|z|$ for each of these stars can also be estimated, but more relevant quantities are the $|z|$
of the center of the cloud's head and the $|z|$ of the tips of the tails.
Using the star HD 215733, taking into account the difference between its Galactic latitude ($b = -36.4^{\rm{o}}$)
and that of the center of the head ($b = -36^{\rm{o}}$), and
making the approximation that the cloud is oriented perpendicular to the Galactic midplane
yields
upper limits of $\sim1700$~pc
and $2050 \pm 530$~pc on the $|z|$ of the cloud's head.
Estimating the $|z|$ of the cloud's head from the distance to the star PG~0039+049 and the Galactic
latitudes of the head and the star is a
less justifiable exercise, owing to the greater angle between PG~0039+049 and the cloud's head, but yields
a  much smaller upper limit on the $|z|$ of the head's center:  $410 \pm 160$~pc.
%
%

The $|z|$ of the tips of the tails can also be estimated.
The tips of the tails extend to a slightly higher $|b|$ than the location of the star PG~0039+049.
Both tips extend to $b \sim -62^{\rm{o}}$,
while the star is at $b=-57.64^{\rm{o}}$.
Taking this difference into account and making the aforementioned approximation about the cloud's orientation
yield an upper limit on the $|z|$ of the tail tips of $1060 \pm 400$~pc.
For completeness, we also
present the upper limits on the $|z|$ of the tail tips
calculated from the distance to
the star HD 215722 and the same assumption about the cloud's orientation, although
the resulting constraints ($|z| \leq \sim4380$~pc and $5280 \pm 1360$~pc) are very loose.

Wakker (2001) estimated the mass of the entire \ppa\ structure to be 
$\leq 5 \times 10^4$ M$_\odot$ by integrating the \hone\ signal in
the Leiden-Dwingeloo Survey data (Hartmann \& Burton 1997) {{across the $V_{\rm{LSR}} = -85$ to $-45$~km~s$^{-1}$
    velocity range, assuming that the cloud is $\leq 1050 \pm 400$~pc from Earth,
    scaling the HI mass by a factor of 1.39 in order to account for the estimated He content of the gas,
    and scaling by a factor of 1.2 in order to account for the estimated \hii\ content of the gas.
Removing both of those scalings yields an \hone\ mass of $\leq 3 \times 10^4$~M$_\odot$.
Later, Fukui et al. (2021) followed up with archival GALFA-\hone\ data (Peek et al. 2011).   They
considered only the head of the cloud, 
estimating its mass of \hone\ gas to be $7 \times 10^3 (d/{\rm{1kpc}})^2$ M{$_\odot$}.
Their estimate does not include He or \hii.
For the same assumed distance as used in Wakker (2001), this equates to $7700$~M$_\odot$,
which is considerably less than the
\hone\ mass of
the whole cloud.
A substantial fraction of the cloud gas may be in the ionized phase that is not observed in the 21~cm
observations.
According to Fukui et al (2021), the \hone\ column density on sight lines through peaks within the head region
is $2 \times 10^{20}$~cm$^{-2}$.   When simulating the cloud, the current column density is a useful starting point
in the search for good initial cloud parameters.


The head is centered at $\ell = 86^{\rm{o}}$, $b = -36^{\rm{o}}$ in \hone\ maps.
Its radius is roughly 3$^{\rm{o}}$.
Its shape is quite asymmetric,
and in general, the denser part of the head is elongated in the same direction as the tails are:  
the low $\ell$, small negative $b$ to high $\ell$, larger negative $b$ direction, 
which we will call the northwest to southeast direction in Galactic coordinates
(see figures in Wakker (2001) and in Fukui et al. (2021).
%

From the Earth's point of view, the \ppa\ travels at negative intermediate velocities.
This is seen in the velocity map,
Figure~\ref{fig:fukui_hi_map}(b)
in this article, which
has been adopted from Figure~4 in Fukui et al. (2021).   It can also be seen in the velocity
channel map in Fukui et al. (2021), i.e., their Figure 2.
The cloud's head has a typical \los\ velocity with respect to the LSR of around $-50$~km~s$^{-1}$.
Some material moves as fast as ${-70}$~km~s$^{-1}$ and some moves as slow as ${-30}$~km~s$^{-1}$.
There is a gradient across the
head,
such that the flat-sided Galactic northeast portion of the head
moves at more extreme negative velocities than the relatively diffuse Galactic southwest
portion of the head.



The velocity map is also helpful for identifying the two tails, because they have different velocities.
The tails diverge from each other around $\ell = 94^{\rm{o}}$, $b = -43^{\rm{o}}$.
The southwestern strand travels at $V_{\rm{LSR}} \sim -65$ to $-70$~km, while the northeastern strand
travels at $V_{\rm{LSR}} \sim -40$ to $-57$~km.    The line of sight velocity varies non-monotonically along each tail,
as if both strands are wavering.
The southwestern strand makes a straight line on the sky,
while the northeastern strand is curvier and more disjointed.
Considering its extreme head-tail morphology, the \ppa\ is assumed to have traveled in the direction of its long axis.
On the plane of the sky, the long axis is oriented at a $\sim45^{\rm{o}}$ angle to the Galactic midplane.
Thus, it gives the appearance that
the cloud has been moving northward at about the same speed that it has been moving westward.
Any motion perpendicular to those two directions is undetermined.

    We next consider Galactic rotation's effect on the \ppa.   Observations of external spiral galaxies have
    found that their extraplanar gas rotates like the disk does, but at a slightly slower speed.
    In a study of    15 disk galaxies, Marasco et al. (2019) measured the lag to be approximately -10~km~s$^{-1}$~kpc$^{-1}$.
    Applying this gradient to our Galaxy, and considering the \ppa's nearness to the midplane,
    yields a small estimated lag of $\lesssim$ 10~km~s$^{-1}$.
    Thus, in the region of the \ppa, the thick disk interstellar matter (ISM) should be moving at a couple of hundred km~s$^{-1}$.   Its direction of
    motion is toward $\ell = 90^{\rm{o}}$.    The \ppa\ 
    lies across its path, from $\ell = 86^{\rm{o}}$ to $b = 115^{\rm{o}}$ and $125^{\rm{o}}$.
    If the cloud's long axis is roughly perpendicular to the line of sight to it (as Fukui et al. (2021) suspect),
    then the \ppa\ is being broadsided by the movement of the ISM's thick disk.
    In contrast, if the long axis of the cloud had been parallel to the direction of flow, then Galactic rotation could have been
    suspected of stretching out the cloud.   But, it does not.
    Nor is the gradient in the ISM's angular velocity
    large enough to suspect it of having stretched the \ppa\ into the long object we see today.
The LSR velocities of the cloud are negative, indicating that the cloud is currently moving
against the direction of Galactic rotation;  the \ppa\ is moving downstream slower than the disk is.
Fukui et al. (2021) observed the \hone\ in the vicinity of the head of the cloud.   Aside from a {\it{bridge}} of
material that has been hit by the cloud's head, the background ISM has a \los\ velocity component
with respect to the LRS of approximately
-10 to approximately 0~km~s$^{-1}$.

Lastly, using the same \hone\ dataset as was used by Fukui et al. (2021), we created a map of the velocity dispersion in the
head of the cloud,
Figure~\ref{fig:fukui_hi_map}(c).
The greatest velocity dispersion (i.e., $\sim{8}$ km s$^{-1}$) is in
the fast-moving ridge of gas on the northeastern side of the head,
while the least dispersion (i.e., $\sim{1}$ km s$^{-1}$) is in the
slower, more diffuse southwestern extension of the head.
Between these two extremes is the main portion of the cloud's head, which has a velocity
dispersion ranging from
$\sim3$ to $\sim7$ km~s$^{-1}$.

\section{Simulations}
\label{sect:simulations}

We use version 4.3 of the FLASH simulational framework
(Fryxell, 2000)
in order to 
model the hydrodynamics as the cloud moves through the Galactic thick and thin disks.
%
FLASH has already been used by several groups to simulate HVCs
in a wide variety of circumstances,
e.g., Orlando etal. (2003); Kwak etal. (2011);  Pl\"{o}ckinger \& Hensler 2012; Galyardt \& Shelton (2016); 
Gritton, Shelton \& Galyardt (2017); Sander \& Hensler (2019); Sander \& Hensler (2020).

The hydrodynamics module in FLASH tracks gas flows, including those leading to
Kelvin-Helmholtz instabilities, Rayleigh-Taylor instabilities, and the resulting turbulent diffusion.
It is also models shocks.
Thermal conduction was not modeled in our simulations.   Between turbulent diffusion and thermal conduction, the former
is substantially more efficient at transporting heat according to 
de Avillez \& Breitschwerdt (2007)
and so is the more important process.
However,
Armillotta et al. (2017)
pointed out that thermal conduction
damps hydrodynamic instabilities, and consequently 
can affect the erosion and spatial distribution of
cold cloud material.
Their point is based on models of 
cold ($T = 10^4$~K) clouds traveling through hot ($T = 2 \times 10^6$~K), rarefied ($n = 10^{-4}$~cm$^{-3}$)
halo-circumgalactic media in
two-dimensional hydrodynamic simulations, some of which employed thermal conduction and some of which did not.
Visual images from sample cases having a cloud speed of 100~km~s$^{-1}$
were presented in Figure 4 of
Armillotta et al. (2017).
In the nonthermally conductive simulation, the cooler, denser gas
had become distributed into a fine filigree, like that expected from
hydrodynamical instabilities.   In contrast, the image of the thermally conductive cloud
is far more muted and contains far less fine-scale structure.
Our case, however, is quite different from the 
case simulated by
Armillotta et al. (2017).
In our case,
the combination of ambient conditions and cloud speed results in a bow
shock that greatly reduces the shear speed between the cloud and the ambient gas.   As a result, there are no
small-scale Kelvin-Helmholtz instabilities evident in the images of our simulations.
Since there is no network of strong, small scale
temperature fluctuations for thermal conduction to wash out,
there is no need to model thermal conduction in our case.

Our simulations model three-dimensional space.   The domain
is gridded 
in Cartesian coordinates and adaptively refined using PARAMESH
(MacNeice et al. 2000).
We initialized the domain to model a cloud surrounded by Galactic thick
disk gas and acting under the influence of gravity.
The thick disk's gas density and temperature
as functions of height above the plane were set to be in approximate hydrostatic balance with the Galaxy's
gravity, whose gravitational potential was determined from the
Galactic mass distribution
and the methods described in
Galyardt \& Shelton (2016).
In order for the gas pressure gradient to balance the Galaxy's gravitational pull,
the gas temperature varies with height above the midplane.
It is $\sim10^3$~K at the Galactic midplane and is higher in the thick disk.
The model temperature exceeds $10^5$~K farthest from the midplane.
Considering that we model gravity and aim to maintain 
hydrostatic balance in the background gas, we cannot allow radiative cooling
in these simulations; if cooling were allowed, the background temperature and
pressure would decrease over time, causing the background material to collapse.
The modeled hydrostatic equilibrium is not completely stable, however.
As a result, a small pressure wave moves vertically through the domain.
However, this pressure wave does not appear to have any significant effects on the IVC during the development of
the tails, through the time period when the tails best mimic those of the \ppa, and through the remainder of the
simulations.  

\begin{deluxetable}{ccc} 		
   \tablecolumns{3}
   \tablecaption{Initial Simulational Parameters}
   \tablehead{		
      \colhead{Parameter} &\colhead{IVC 1} & \colhead{IVC 2} }
\startdata
 $z$                      & -1100 pc                    & -2000 pc   \\
 radius 	          & 43.7 pc                     & 87.3 pc   \\ 
%
%
 hydrogen number density  & 0.89 cm\textsuperscript{-3} & 0.45 cm\textsuperscript{-3}     \\
%
%
%
hydrogen column density (center) & 2.4$\times$10\textsuperscript{20} cm\textsuperscript{-2} & 2.4$\times$10\textsuperscript{20}  cm\textsuperscript{-2}       \\ 
mass               & $\sim{8900}$ M\textsubscript{$\odot$} & $\sim{3.6}$$\times$10\textsuperscript{4} M\textsubscript{$\odot$}    \\
 velocity in $\hat{x}$ direction & 70 km s\textsuperscript{-1} & 70 km s\textsuperscript{-1}   \\
velocity in $\hat{z}$ direction & 70 km s\textsuperscript{-1} & 70 km s\textsuperscript{-1}   \\
 \enddata

\label{table:parameters}
\end{deluxetable}

The domain is designed such that the domain's $xy$ plane is parallel to the Galactic midplane and the $\hat{z}$ direction
runs perpendicular to the midplane.
The $\hat{x}$ direction is perpendicular to the line of sight's projection onto the midplane 
and the $\hat{y}$ direction is parallel to the line of sight's projection onto the midplane.
The \ppa\ has a negative Galactic latitude and so a negative value of $z$.
The Galactic midplane is placed in the upper fifth of the domain.
When describing images made from the distribution of the simulated material in the $xz$ plane,
the convention used to describe directions is
analagous to that used when discussing the observations.
I.e., east (low values of $x$) is to the left, west (higher values of $x$)
is to the right, north (positive values of $z$) is up, and south (negative values of $z$) is down.

At the beginning of each simulation, a spherical cloud is initialized in the lower left
corner of the domain.
%
%
The temperature and density in the cloud are set such that the cloud is initially in pressure balance
with the ambient medium
and
the cloud's radial temperature and density distributions make
graduated transitions
from the center of the cloud to the outer edge where the cloud meets the ambient
gas.   These distributions are described in
Galyardt \& Shelton (2016).
%
%
The cloud is given an initial overall velocity of
$\sim{100}$~km~s$^{-1}$ directed
at a 45$^{\rm{o}}$ angle toward the Galactic midplane.
This angle
is roughly consistent with the observed angle of the \ppa's long axis, which is the angle at which the
\ppa\ is thought to have moved. 
The initial velocity is entirely in the $xz$ plane.
We developed two simulational models, one at a nearer distance and one at a farther distance,
whose morphology and whose head's \hone\ 21~cm intensity and velocity generally agree with the observations.
%
They are IVC~1 and IVC~2 (Parker 2019).
%
%

For Simulation IVC~1,
%
we start the simulation with the cloud located 1100~pc below the midplane and
set the lower $z$ boundary of the domain  $\sim1200$~pc below the Galactic midplane.
This placement provides some
space around the cloud at the beginning of the simulation.
We set the upper $z$ boundary $\sim400$~pc above the Galactic midplane so that the future collision between
the cloud and the Galactic disk can be modeled.
In this simulation,
the domain size is 1088 pc in the $x$ direction, 128 pc in the $y$ direction, and 1600 pc
in the $z$ direction.
When the grid is fully refined,
the maximum number of cells is 544 $\times$ 64 $\times$ 800 cells and the cell sizes are 2 pc $\times$ 2 pc $\times$ 2 pc.
%
In Section~\ref{sect:results}, we apply the constraint that the center of the head of the simulated cloud must have
a latitude of $-36^{\rm{o}}$ at the moment when the model most resembles the \ppa.
This constraint places the IVC~1 cloud 810~pc from Earth at that time.

In order for the IVC~2 simulation to have the same minimum cell size as the IVC~1 simulation, the number of cells in the
domain is roughly proportional to the cube of the height of the domain, which is constrained by the initial location
of the model cloud.  Given computational limitations, we set the cloud 2000~pc below the midplane,
the lower $z$ boundary of the domain $\sim2100$~pc below the midplane, and the upper $z$ boundary
$\sim500$ above the midplane.
The domain size is 2336 pc in the $x$-direction, 192 pc in the $y$-direction, and
2592 pc in the $z$-direction.
The number of cells in the domain is 1168 $\times$ 96 $\times$ 1296 cells
and the minimum cell size is 2 pc $\times$ 2 pc $\times$ 2 pc after refinement.
As shown in Section~\ref{sect:results},
applying the constraint that the center of the head of the simulated cloud must be located at $b = -36^{\rm{o}}$
at the time when the cloud most resembles the \ppa\
constrains the distance to the head of the simulated cloud to be 1530~pc at that time.

The cloud's initial size and hydrogen number density were chosen as a result of
trial and error with preliminary simulations and from 
the current size and hydrogen number density of the head of the \ppa.
In Simulation IVC~1,
the initial cloud radius is 43.7 pc and 
%
%
%
the initial cloud hydrogen number density is 0.89 cm\textsuperscript{-3}.
In Simulation IVC~2, 
the initial cloud radius is 87.3 pc and the initial
cloud hydrogen number density is 0.45 cm\textsuperscript{-3}.
%
%
Table~1 lists these and other initial
values
for the simulations.   The quoted hydrogen density is the number of hydrogens per cm$^{3}$ in the central part of the cloud at the
zeroth epoch.  The simulations include helium, which contributes to the overall density of the material.

\vspace{2cm}

\section{Results}
\label{sect:results}

\subsection{Development of the Cloud's Tail and Head}
\label{sect:morphology}

      Figures~\ref{fig:ivc1_density_slices} and \ref{fig:ivc2_density_slices} show several epochs in
      the evolution of Simulations IVC~1 and IVC~2.
In each series of snapshots, the cloud moves toward the Galactic disk at an oblique angle, deforms, grows a tail,
collides with the Galactic disk, and disrupts.
The final panels in each figure show that the disk is also disturbed by the collision.



\begin{figure}[H]

\includegraphics[trim=170 200 200 200, width=150pt, height=250pt]{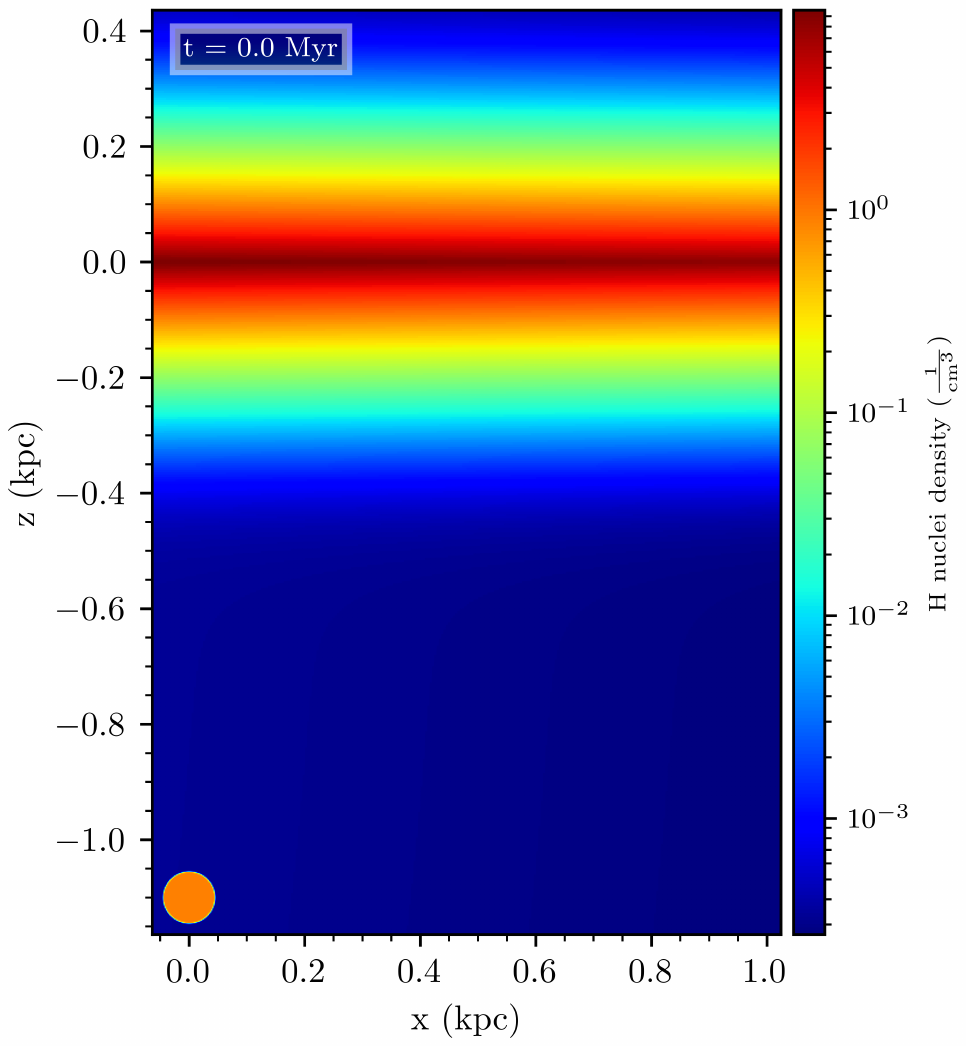}{(a)}
\includegraphics[trim=170 200 200 200, width=150pt, height=250pt]{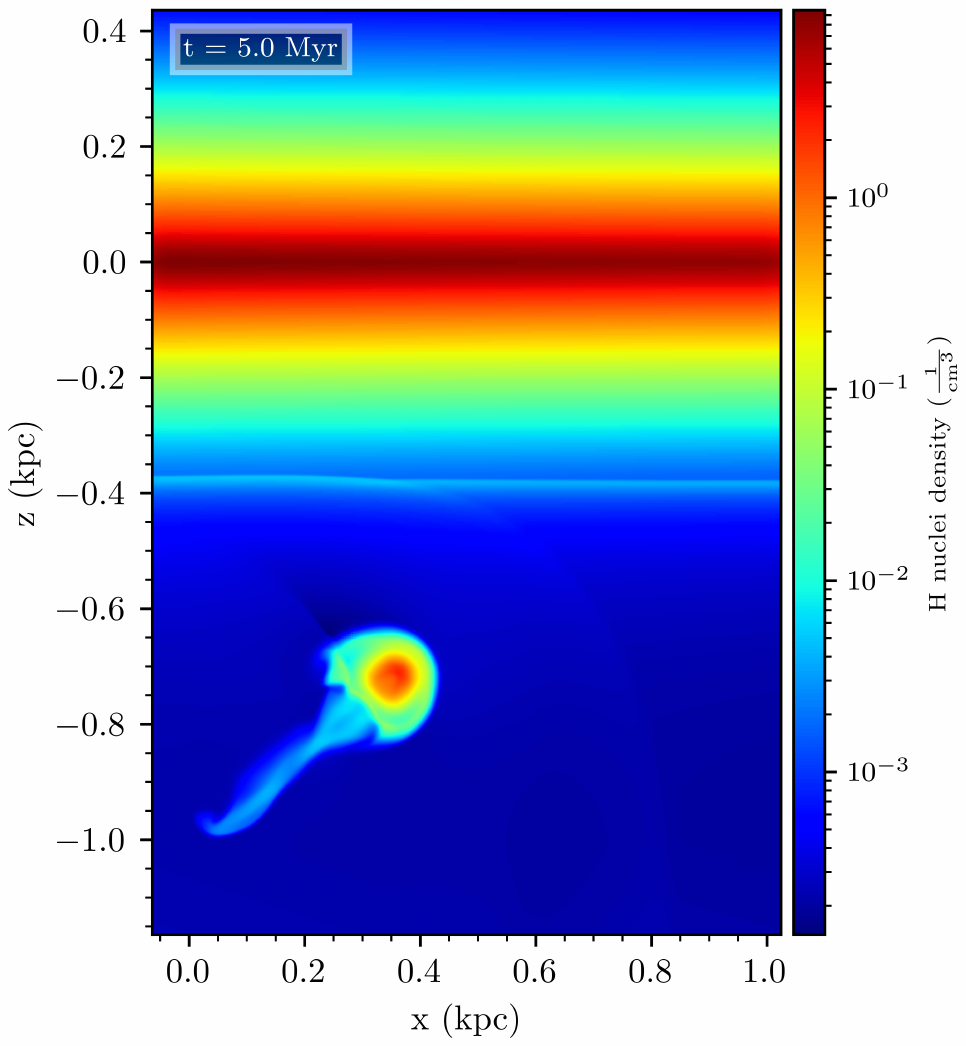}{(b)}
\includegraphics[trim=170 200 200 200, width=150pt, height=250pt]{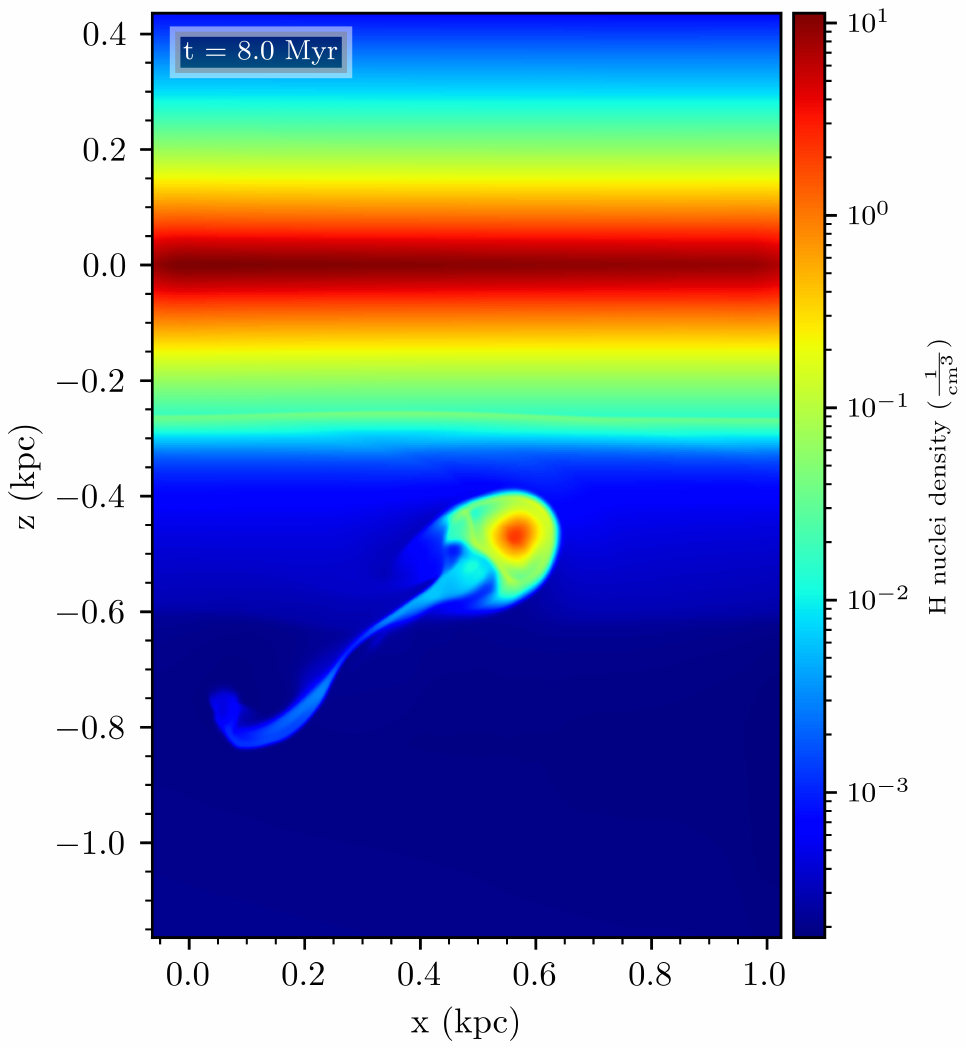}{(c)}

\includegraphics[trim=170 200 200 200, width=150pt, height=250pt]{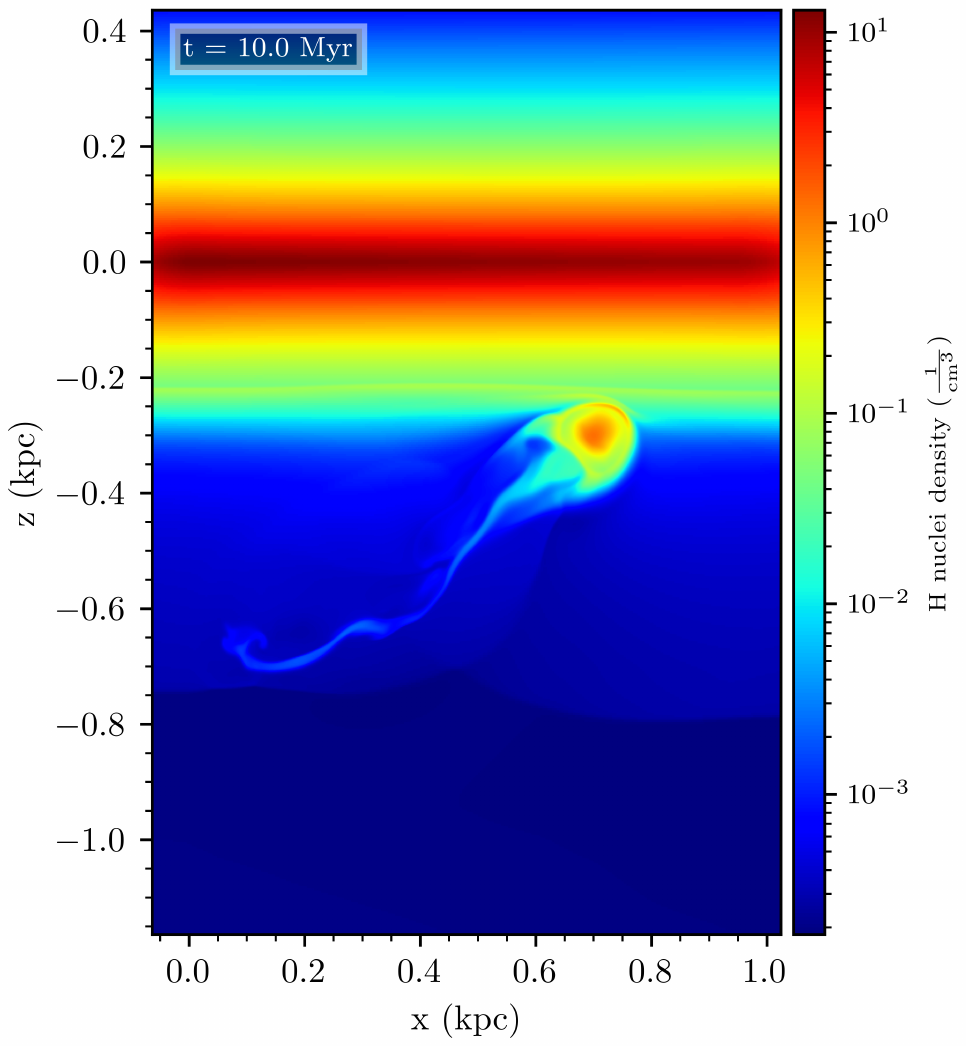}{(d)}
\includegraphics[trim=170 200 200 200, width=150pt, height=250pt]{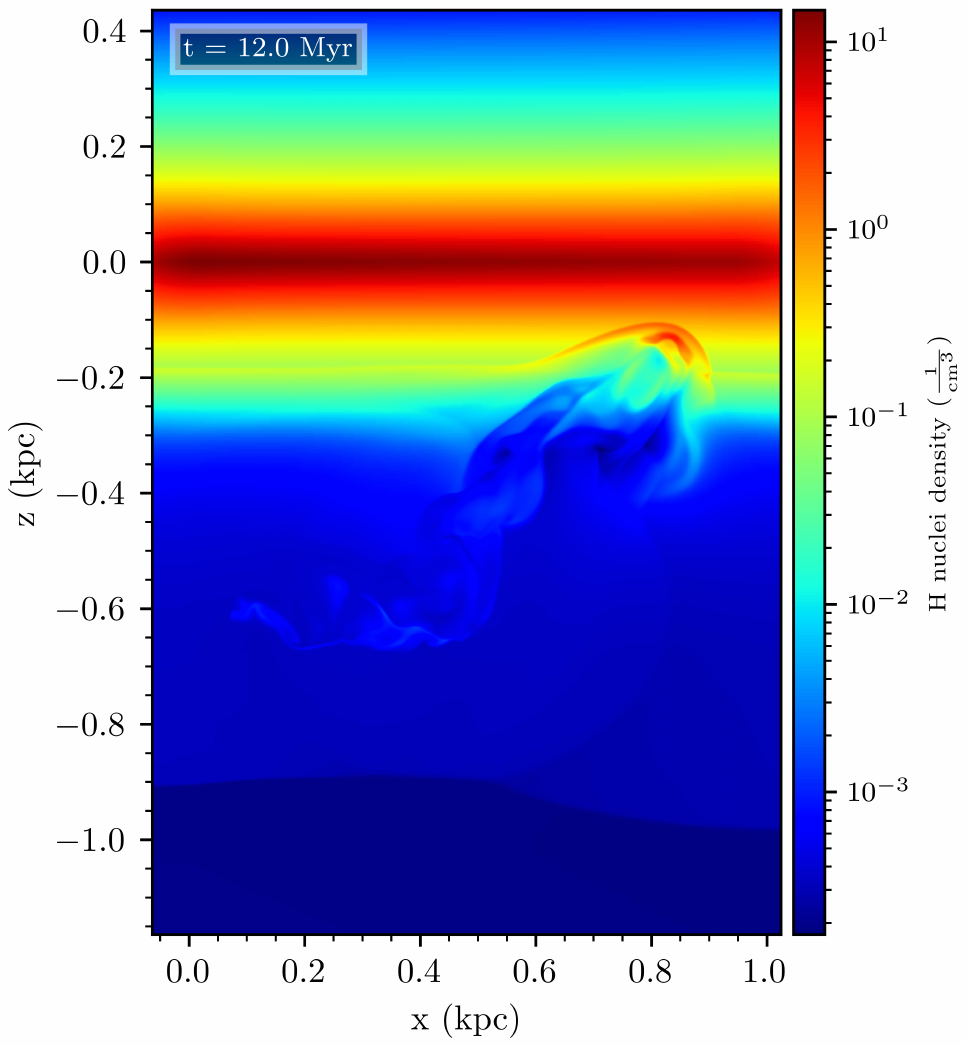}{(e)}
\includegraphics[trim=170 200 200 200, width=150pt, height=250pt]{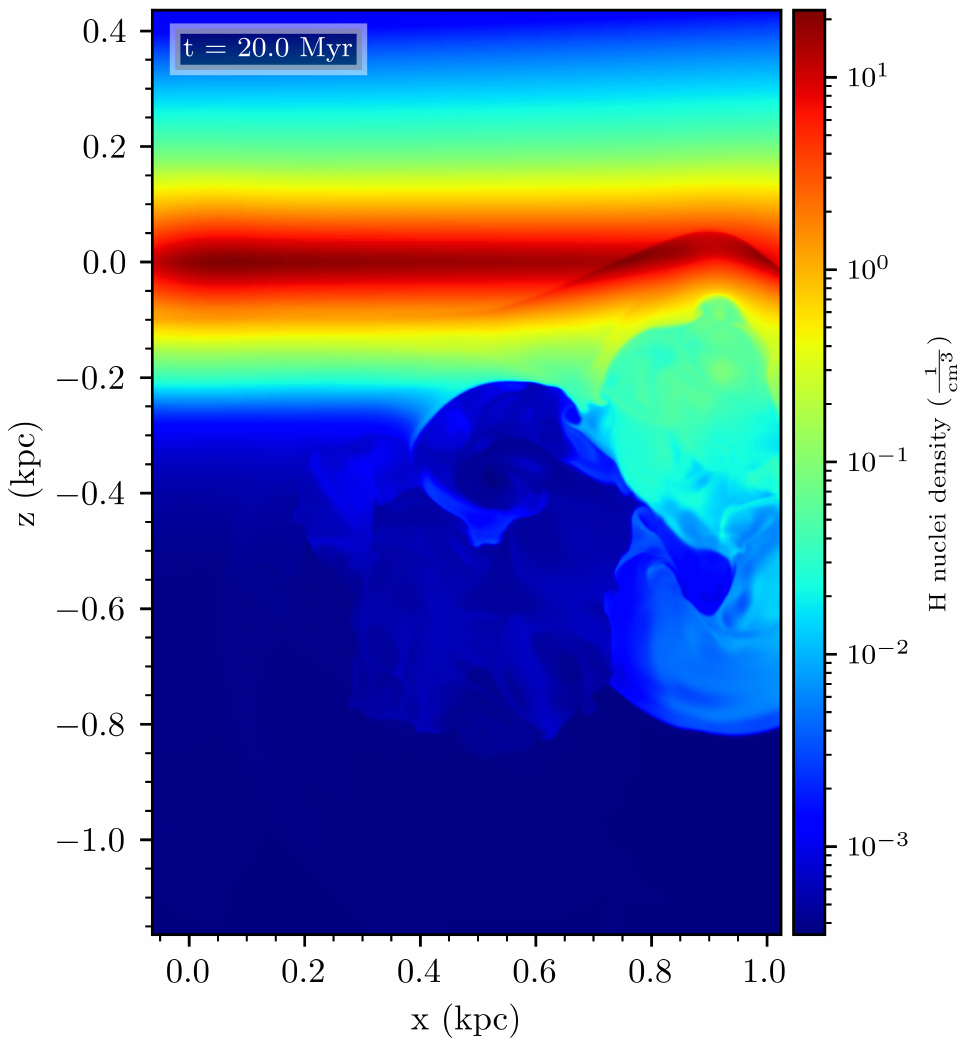}{(f)}



%
  \caption{Evolution of IVC~1 over 20~Myr of simulated time.
    The panels represent material in vertical slices through the domain at the following times after the
    beginning of the simulation:
   {\it{a}}~$= 0$~Myr,
    {\it{b}}~$= 5.0$~Myr, {\it{c}}~$= 8.0$~Myr, {\it{d}}~$= 10.0$~Myr, {\it{e}}~$= 12.0$~Myr, 
    and {\it{f}}~$= 20.0$~Myr.
    Color is keyed to hydrogen number density
    hence the cloud initially appears in orange, the Galactic midplane appears in maroon, and the thin and thick
    disks appear in a range of colors.
    As shown in this series of snapshots, the cloud moves toward the Galactic disk, deforms, grows a tail that is oriented along
    the direction of motion, and eventually begins to disrupt as it interacts with the Galactic disk gas.
    The evolutionary stage shown in panel (c) is most similar to that of the currently observed \ppa.
}
    \label{fig:ivc1_density_slices}
\end{figure}

\begin{figure*}[htb!]		 
\includegraphics[trim=170 200 200 200, width=125pt, height=200pt]{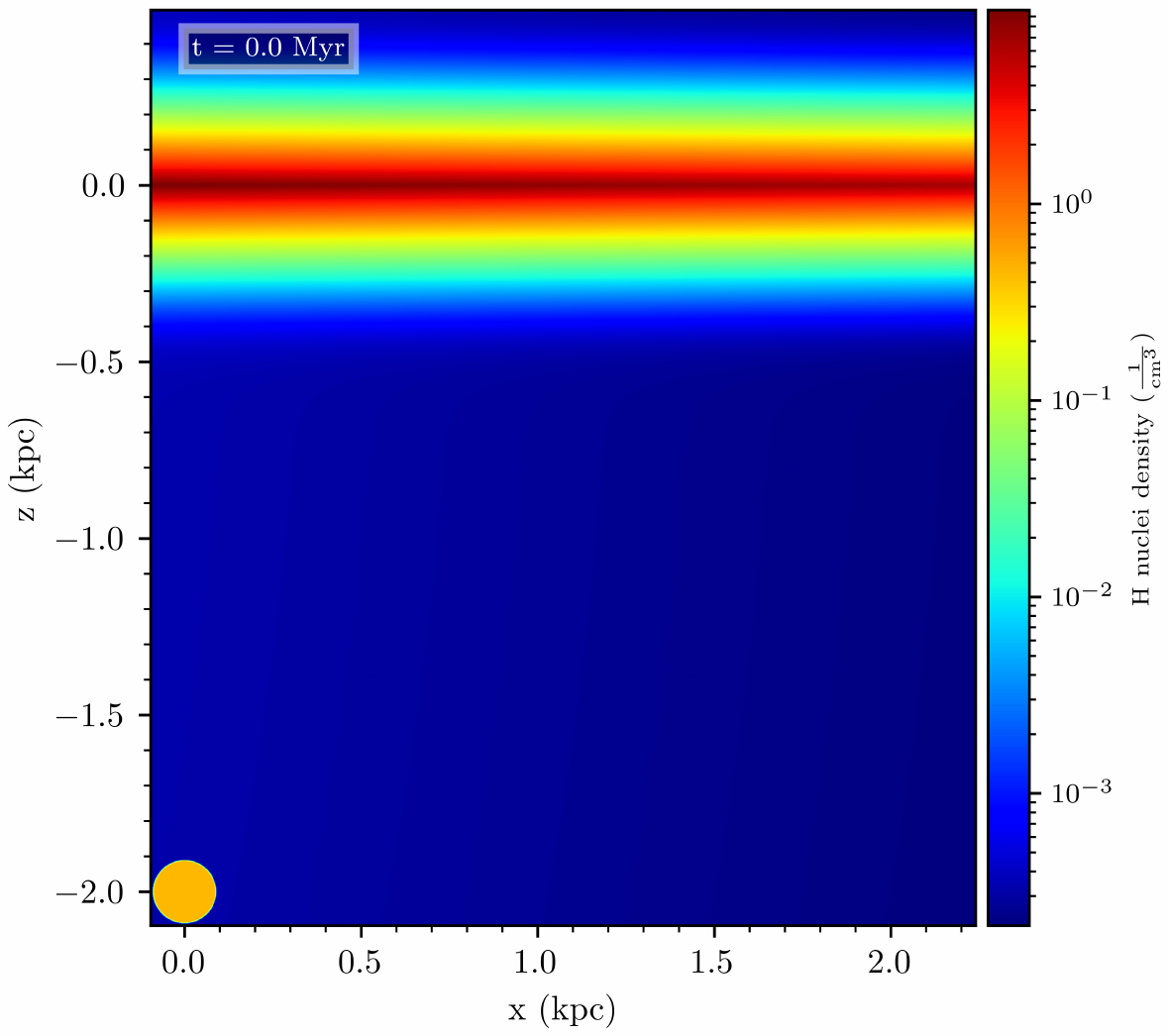}{(a)}
\includegraphics[trim=170 200 200 200, width=120pt, height=200pt]{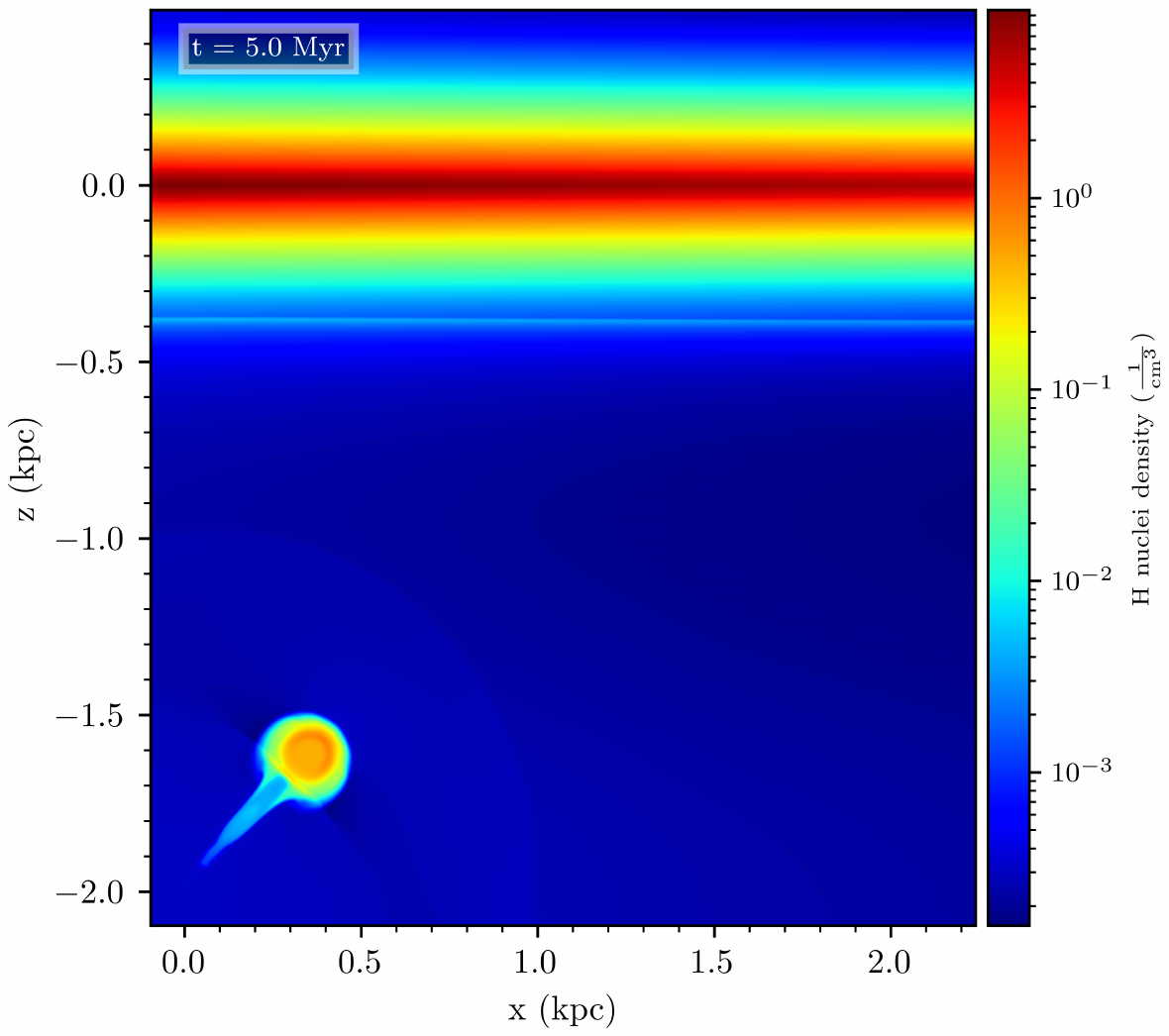}{(b)}
\includegraphics[trim=170 200 200 200, width=115pt, height=200pt]{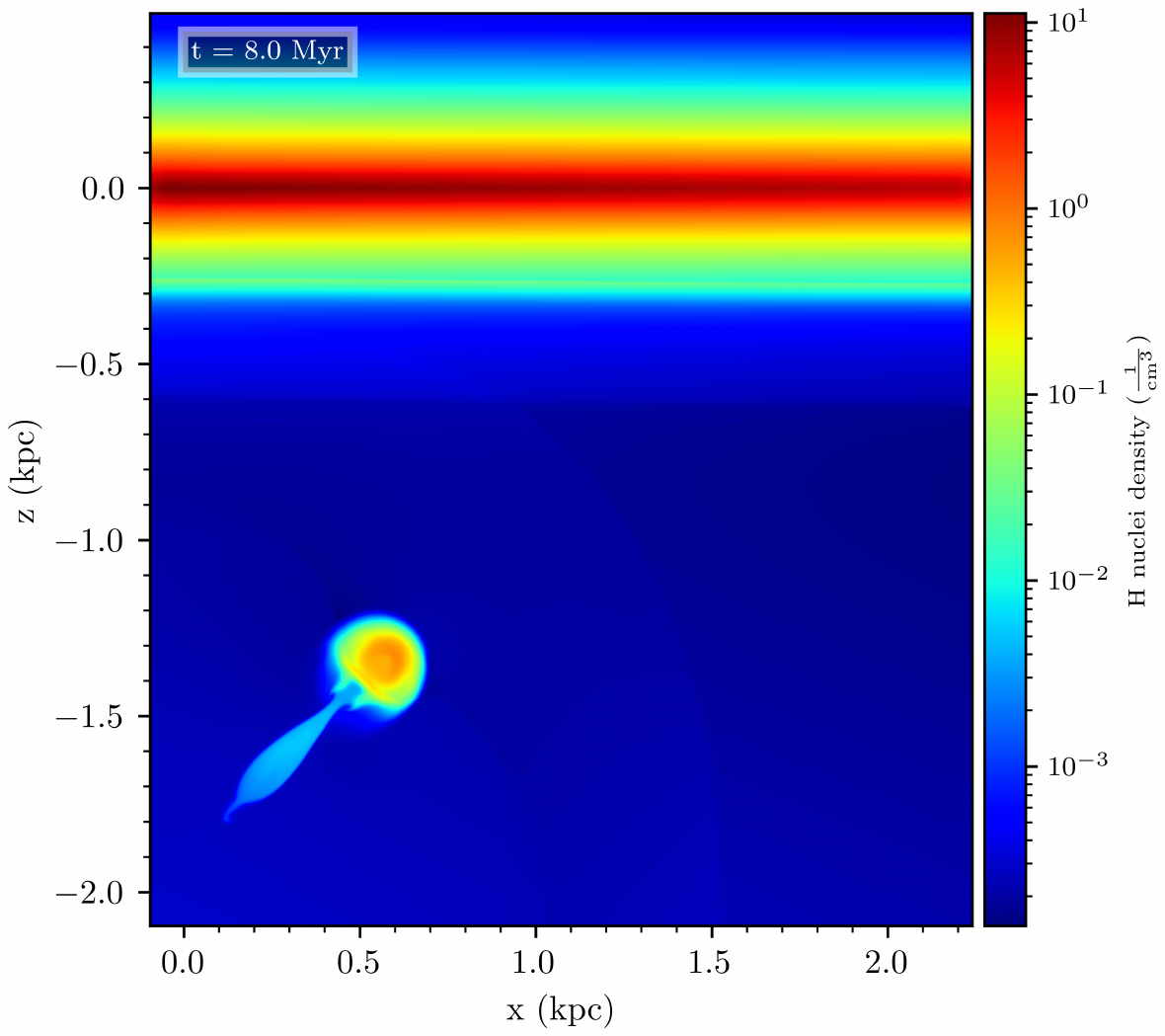}{(c)}

\includegraphics[trim=170 200 200 200, width=125pt, height=200pt]{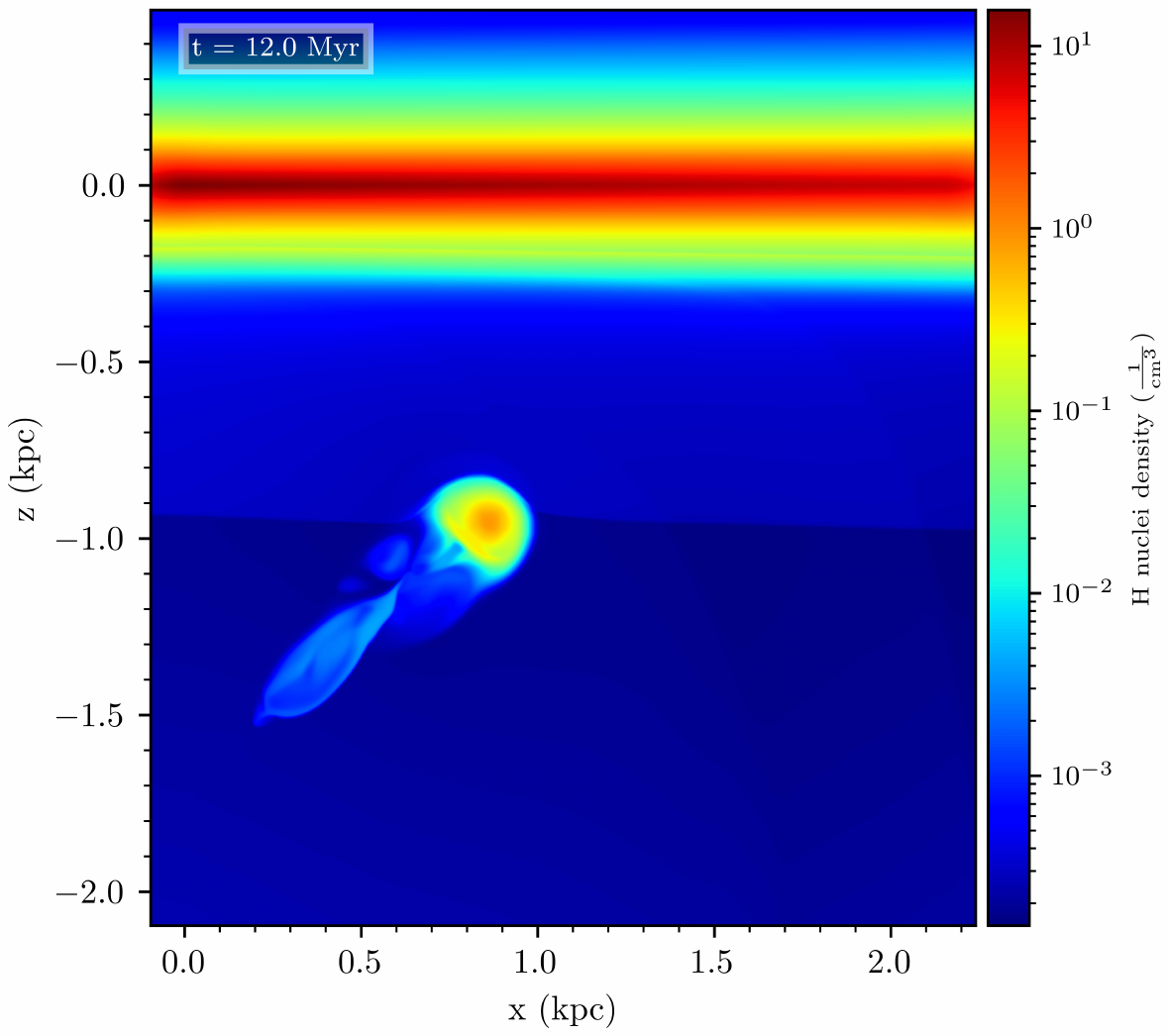}{(d)}
\includegraphics[trim=170 200 200 200, width=120pt, height=200pt]{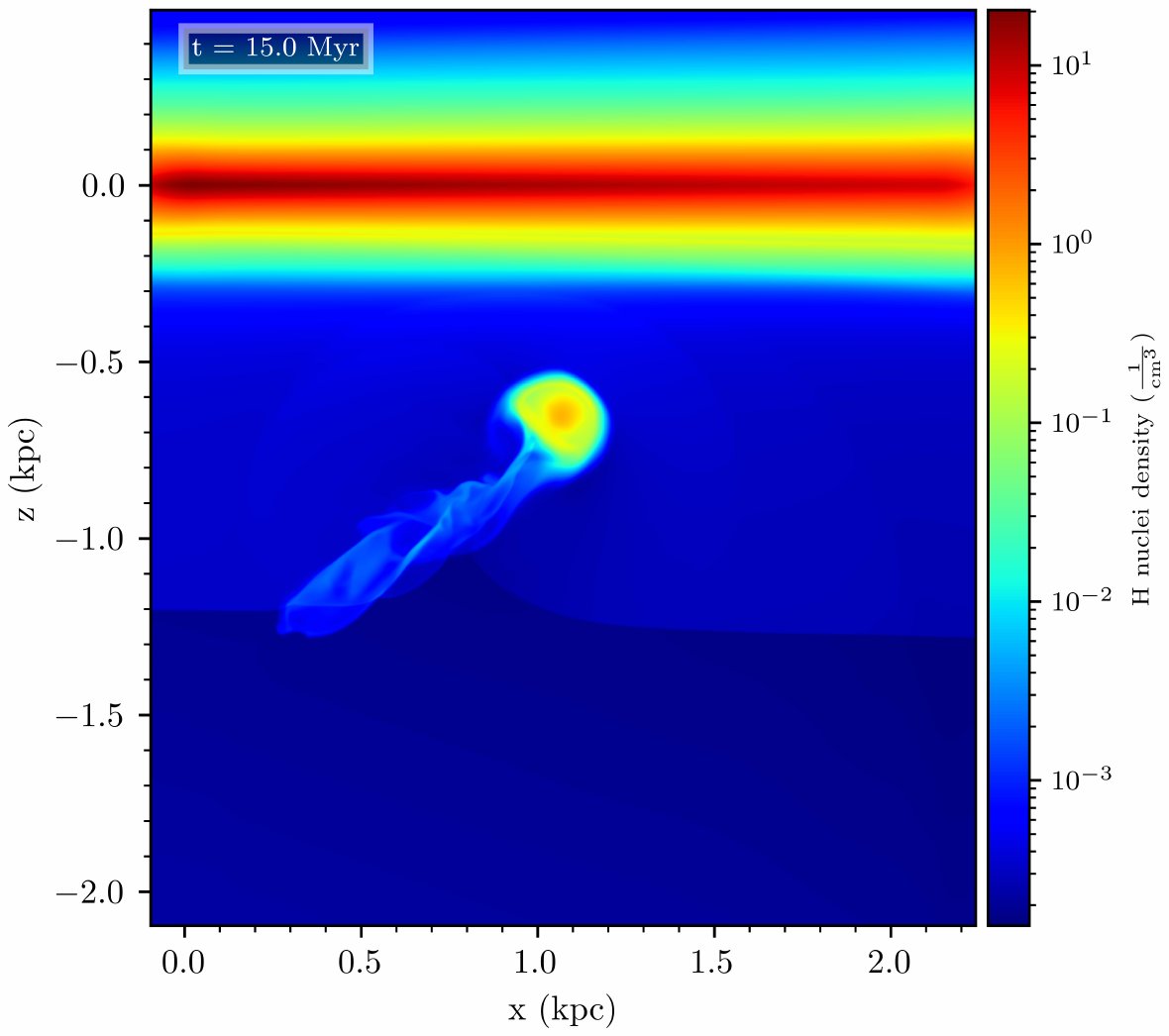}{(e)}
\includegraphics[trim=170 200 200 200, width=115pt, height=200pt]{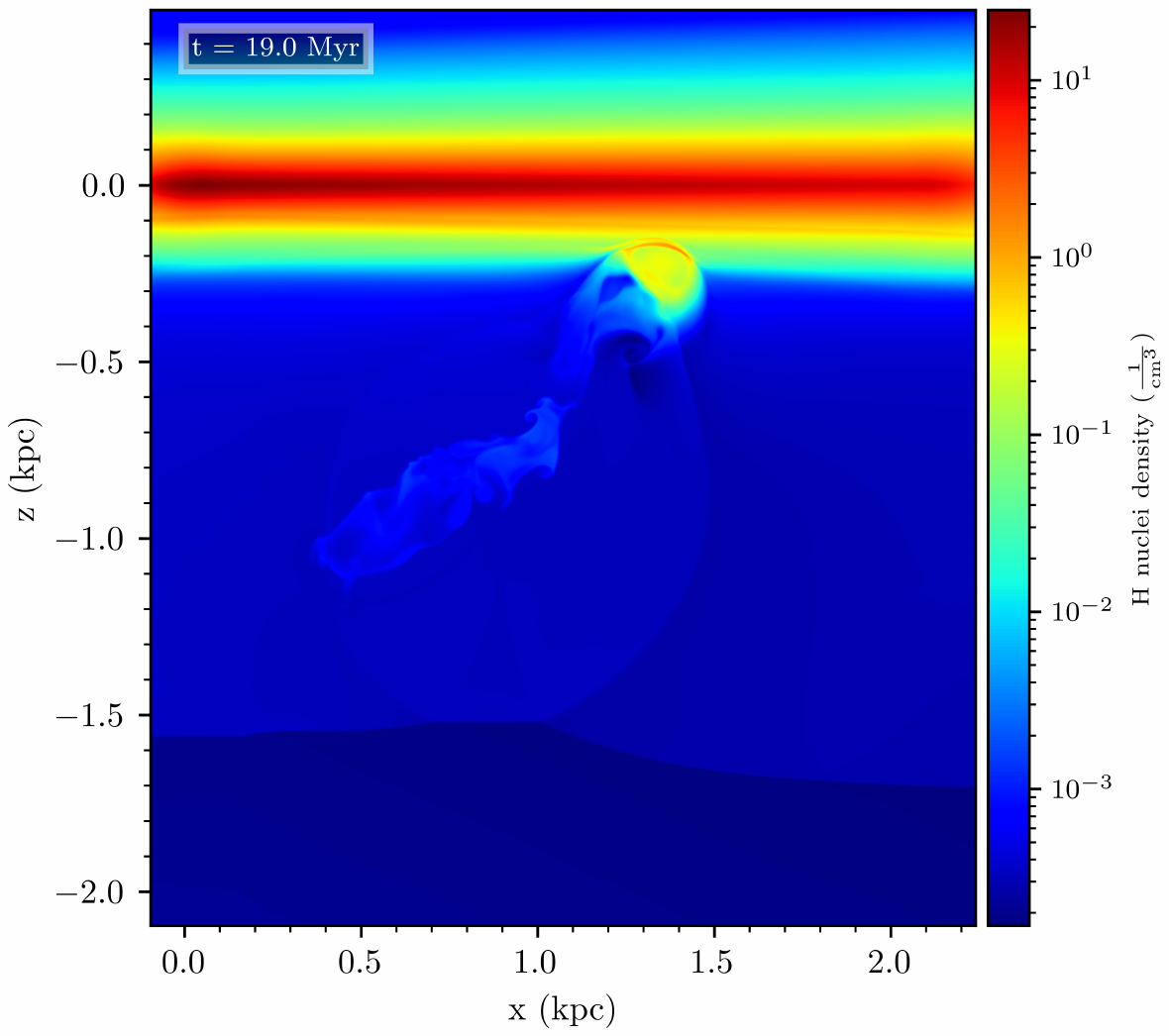}{(f)}

\includegraphics[trim=170 200 200 200, width=125pt, height=200pt]{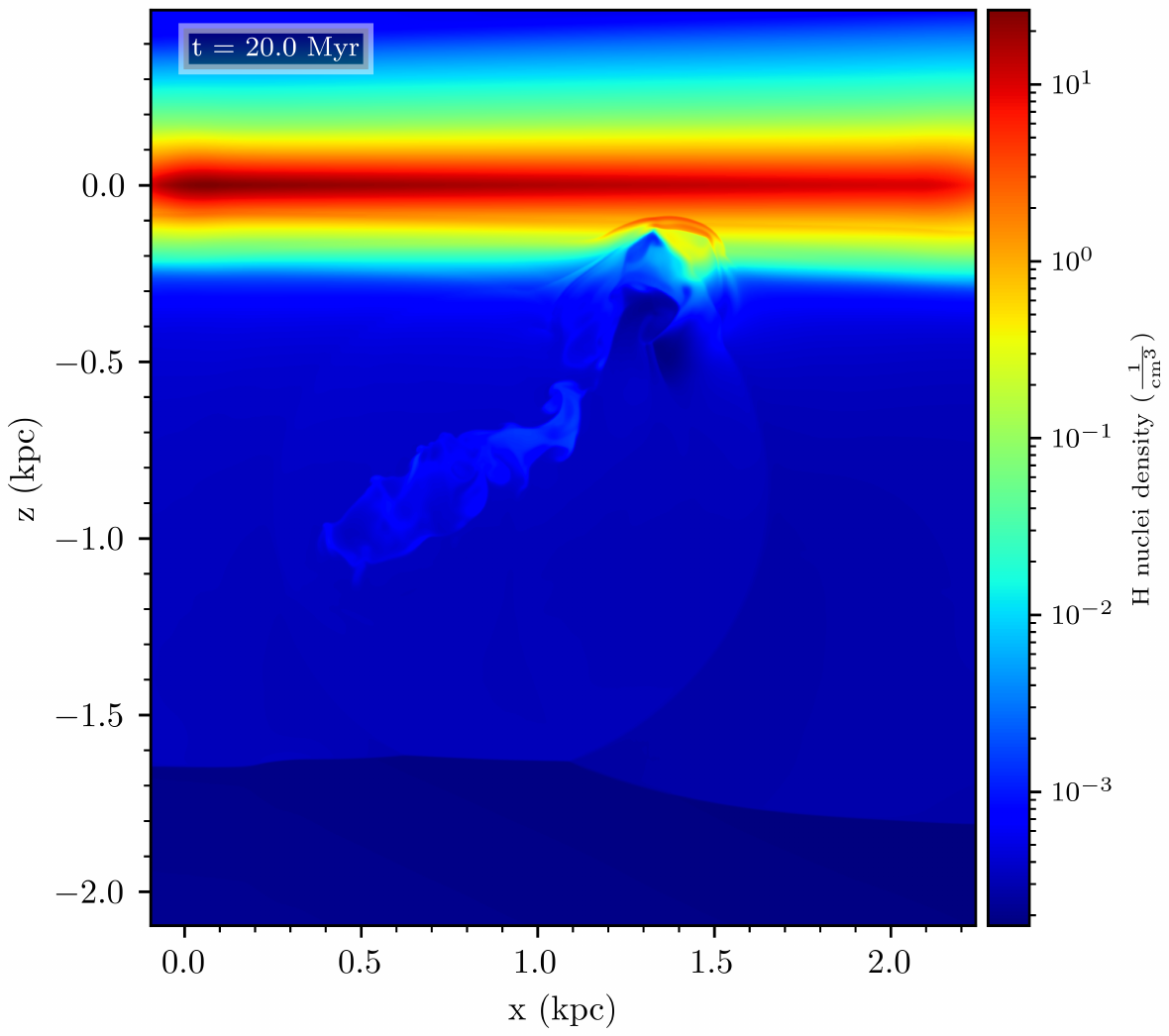}{(g)}
\includegraphics[trim=170 200 200 200, width=120pt, height=200pt]{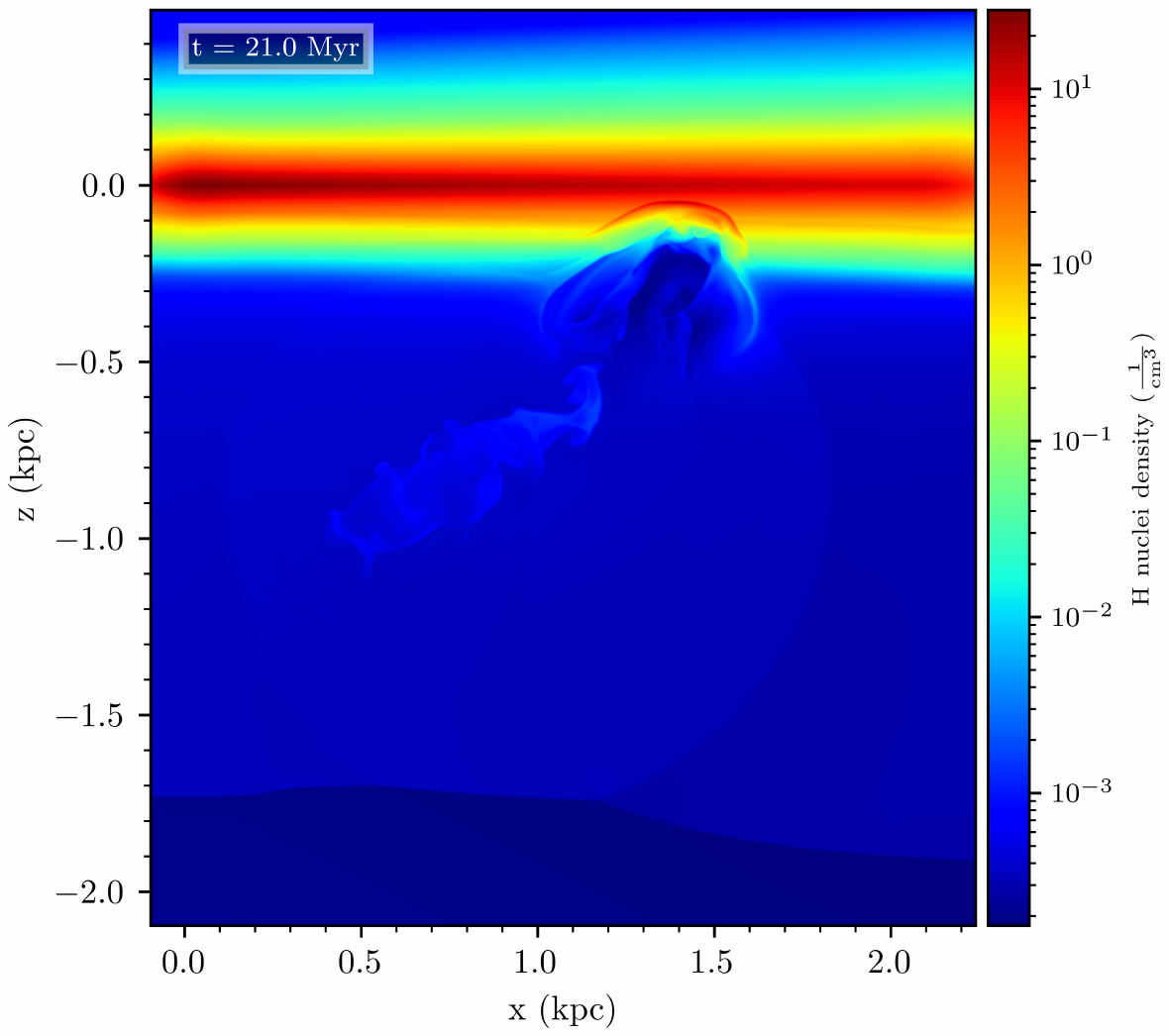}{(h)}
\includegraphics[trim=170 200 200 200, width=115pt, height=200pt]{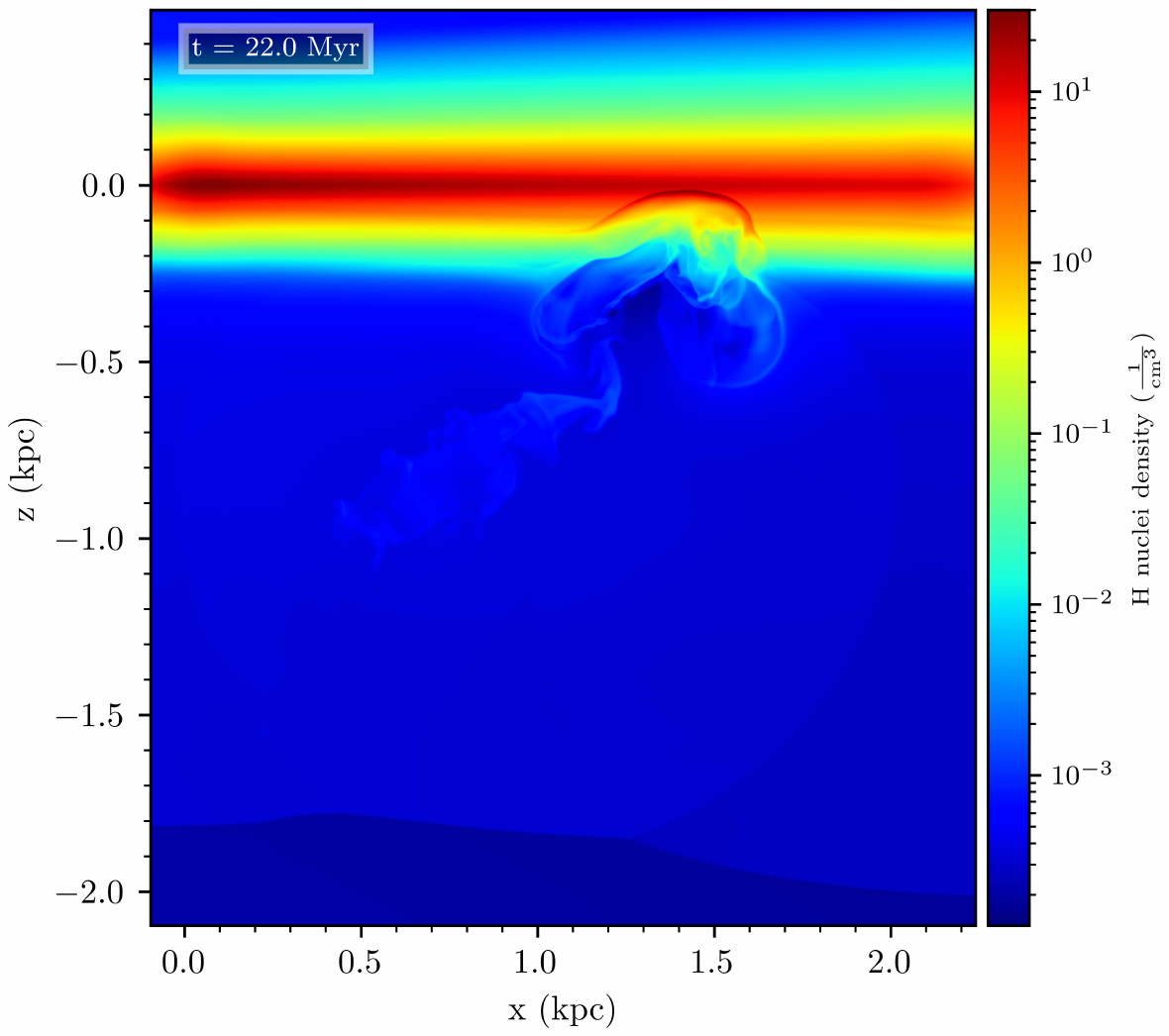}{(i)}

  \caption{Same is in Figure~\ref{fig:ivc1_density_slices}, but for the IVC~2 simulation and evolutionary times:
    (a)~$= 0$~Myr, 
    (b)~$= 5.0$~Myr,
    (c)~$= 8.0$~Myr,
    (d)~$= 12.0$~Myr,
    (e)~$= 15.0$~Myr, 
    (f)~$= 19.0$~Myr,
    (g)~$= 20.0$~Myr,
    (h)~$= 21.0$~Myr,
    and (i) $= 22.0$~Myr.
    The evolutionary stage shown in panel (d) is most similar to that of the currently observed \ppa.
  }
%
\label{fig:ivc2_density_slices}
\end{figure*}


    Both simulated clouds develop long tails that are aligned with the direction of motion.
    Early on, IVC~1's tail is nearly straight, aside from the curl at its end.   Over time,
    the tail stretches into a longer, narrower, more sinuous shape.
    Figure~\ref{fig:ivc1_density_slices}(c) shows a slice through the structure when IVC~1 is 8~Myrs old
    and the tail is several hundred parsecs in length.
%
    This slice transects one of 
    IVC~1's two tail-density enhancements.    The other density enhancement is slightly off axis and so is not
    apparent in this image.
    However, it is revealed by integrating the density
    along the $y$ direction as is done in Sections~\ref{sect:intensity} and \ref{sect:velocity}.
%
    The 8~Myr age in 
    IVC~1's evolution is most like that of the presently observed \ppa, because the tail is longer
    than in previous epochs while the cloud has not yet collided with the denser gas nearer to the Galactic
    midplane, which significantly distorts the cloud.
%


IVC~2's tail also starts with a straight shape, also stretches over time, and also takes on a bifurcated appearance.
The density along a slice through the structure when IVC~2 is most like the \ppa, i.e., when it is 12~Myr old, 
is shown in Figure~\ref{fig:ivc2_density_slices}(d).
By that time, ridges of denser material have developed on the northeast and southwest flanks of IVC~2's trailing gas.
Integrating the density along the $y$ direction reinforces these ridges, creating the appearance of two tails.
Not only do both IVC~1 and IVC~2 appear to have twin tails, but 
twin tails also appear in other preliminary simulations 
and in observed images of the \ppa.

We next consider the heads of the simulated clouds.  They are of interest for comparison with the noticeably asymmetric head of the \ppa.
The northeast edge of the head  of the \ppa\ is flatter, straighter, and more sharply bounded
than the southwest edge, which is rounder, rougher, and more gradiated.
The heads of IVC~1 and IVC~2 also develop sharply bounded, dense edges,
although these edges are more to the north than the northeast and
develop late in the simulated evolutions. 
Consider, for example IVC~2 at 19~Myr shown in Figure~\ref{fig:ivc2_density_slices}(f).
The slight flattening and steepening of the density gradient on the northern side of the head
are due to the cloud's encounter with relatively dense interstellar gas near the Galactic disk.
The leading side of IVC~1 also develops a sharp density gradient.   See
Figure~\ref{fig:ivc1_density_slices} panels (d) and (e) for IVC~1 at 10 and 12~Myr, respectively.

%
Fukui et al. (2021) argue that 
the head of the \ppa\
is colliding with an interstellar cloud.  Their argument is based on a velocity {\it{bridge}} between
that of the cloud and that of the Galactic disk along lines of sight through the head of the cloud.
%
%
They cite similarities between the $-20$ to $-30$~km~s$^{-1}$ velocity {\it{bridge}} along lines of
sight through the head of the \ppa\ and the simulation figures presented in Torii et al. (2017),
which were based on the simulational work done in Takahira et al. (2014).
In accordance with their argument, and 
considering that the northern sides of our simulated cloud heads become compressed and flattened when they encounter
larger ambient densities,
it is reasonable to expect that the northeastern flanks of the simulated clouds would have been correspondingly flattened if
they had encountered similarly dense environmental material, such as another interstellar cloud.
It is also reasonable to speculate that the cloud that developed into the Pegasus-Pisces Arch was initially
asymmetric with a lesser density in its western side than in its eastern side, giving rise to the low density,
western extension we see now in images of the Pegasus-Pisces Arch.

The simulated IVCs develop bow shocks.   They can be seen
upon close inspection of Figures~\ref{fig:ivc1_density_slices}(b) and \ref{fig:ivc2_density_slices}(b).
These bow shocks speed up the ambient material, greatly reducing the velocity contrast
between it and the cloud's head and tails.
The net effect is to protect each simulated cloud from strong
shocks and hydrodynamic instabilities.

\subsection{Simulated \hone\ 21 cm Intensity Maps}
\label{sect:intensity}

We calculated the column densities of intermediate velocity hydrogen on sight lines through the simulated domains.
This was done for IVC~1 at 8~Myr and IVC~2 at 12~Myr and
was done by integrating the densities along lines of sight
running perpendicular to the $xz$ plane in the simulational domains, i.e., parallel to the simulated midplanes.
We then
converted the column densities into \hone\ 21 cm intensities
using the following formula for optically thin gas:
\begin{equation}
I\textsubscript{21\ cm} =
\frac{N_{\rm{HI}}}{1.82\times10^{18} \ \rm{cm\textsuperscript{-2}}}
\ {\rm{K~km~s\textsuperscript{-1}}}.
\end{equation}
The resulting \hone\ 21 cm intensity maps for the two simulated IVCs
are shown in 
Figure~\ref{fig:integrated_intensity}.

\begin{figure*}[htb!]		 


\includegraphics[trim=140 200 150 200, width=205pt, height=300pt]{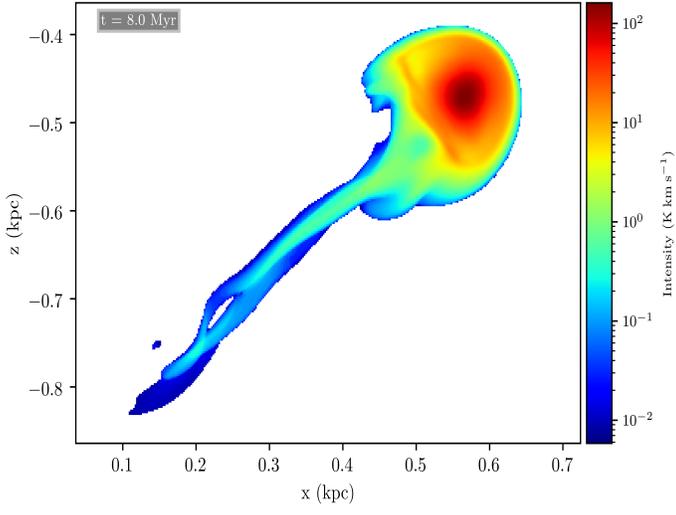}{(a)}\\
\includegraphics[trim=140 200 150 200, width=250pt, height=300pt]{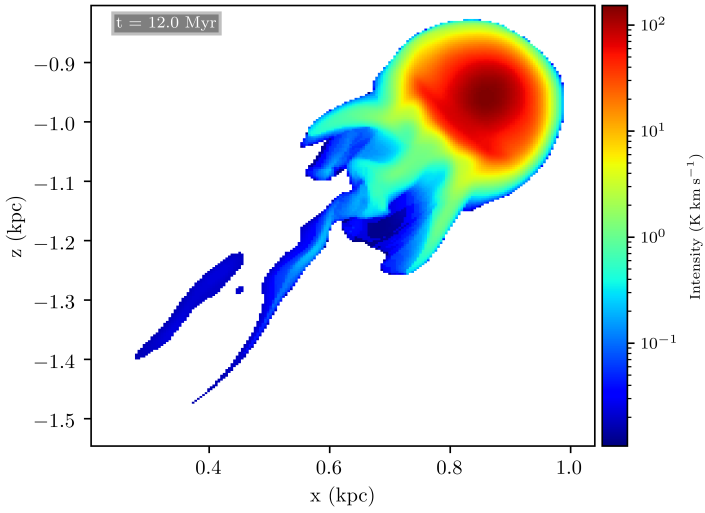}{(b)}

  
  \caption{Maps of simulated intensity for
  (a) IVC 1 at 8.0 Myr and (b) IVC 2 at 12.0 Myr.  The simulated IVCs exhibit cometary morphologies
    with clearly defined heads and tails.
}
\label{fig:integrated_intensity}
\end{figure*}

The \hone\ 21 cm intensities of the central regions of the simulated heads
are around 100~K~km~s$^{-1}$, which is similar
to the
intensity in the central region of the head of the \ppa\ shown in
Figure~\ref{fig:fukui_hi_map}.
%
The maximum simulated intensities ($\sim160$~~K~km~s$^{-1}$) are also similar to those of the \ppa.
However,
the simulated cloud heads have smoother intensity distributions and slower gradients 
than the head of the \ppa, which has a more mottled face and sharper northeast edge.
In addition,
the southwestern extension of the \ppa's head is
brighter and wider than the southwestern portions of the simulated cloud heads.
The head of the \ppa\ probably contains more density inhomogeneities than do the simulated clouds
while the sharp boundary on the northeast side of the \ppa's head 
may be due to a collision with denser ambient material, such as a cloud, as was discussed in
Section~\ref{sect:morphology}.
The intensity plots also reveal that each simulated cloud has a bifurcated tail.
The simulated tails are dimmer than those of the \ppa.
The typical width of each IVC~1 tail is around 43~pc, while that of each IVC~2 tail is around 29~pc.
These widths are many times larger than the 2~pc $\times$ 2~pc $\times$ 2~pc cells.


\subsection{Distances to the Simulated Clouds and the Clouds' Angular Sizes}

The imagined distance between the Earth and the head of either model cloud can be determined if we equate
the Galaxy's midplane with the simulated midplane and equate
the latitude of
the center of the \ppa's head (i.e.,
$b = -36^{\rm{o}}$) with the center of
the luminous part of
the model's head.
First, we consider IVC~1 at 8~Myr, at which time
the luminous part of its head is 475~pc below the Galactic midplane.
Making the approximation that the Earth is in the Galactic midplane
and doing a little trigonometry will determine that the distance between
the head of IVC~1 and Earth is 810~pc.
The $y$ component of this distance is 
650~pc.
Note that in all calculations, we retain significant digits but round the presented numerical results.

The imagined latitude and longitude of the tip of a simulated tail
can also be determined trigonometrically.
%
The first step is to recognize that
the simulations were set up such that the clouds travel in the $xz$ plane.    Therefore,
the $y$ component of the distance
between the Earth and a simulated cloud's head is the same as that between the Earth and each tail.
The tip of the longest tail in IVC~1 at 8~Myr is
%
830~pc below the midplane.
This information, along with the previously determined 
$y$ component of the distance from the Earth to the tail (i.e., 650~pc), yields the tip's latitude,
$b_{t1} = -52^{\rm{o}}$.
In order to determine the longitude of the tip of the longest tail,
we first calculate the
longitudinal span from the tip to the meridian through the center of the bright part of the head, $\Delta \ell$.
We calculate $\Delta \ell$ from the $x$ component of the linear span 
from the tip
to the brightest part of the head,
which is $\Delta x = 450$~pc.
We then make the approximation of treating $\Delta x$ as if it is an arc along a circle that is
located at $b_{t1} = -52^{\rm{o}}$ and that has a radius of 650~pc.
In that case, $\Delta \ell / 360^{\rm{o}}$ can be equated with $\Delta x$ divided by the circumference of
the circle.
This logic yields $\Delta \ell = 39^{\rm{o}}$.
%
Adding a $\Delta \ell$ of $39^{\rm{o}}$ to the longitude of the \ppa's head (i.e., $\ell = 86^{\rm{o}}$) yields an
$\ell_{t1}$ of 125$^{\rm{o}}$.
In summary, if IVC~1 were to be imagined as being located in the sky such that the simulated Galactic midplane
aligns with the real Galactic midplane and 
the center of the bright part of the simulated head is
at $\ell = 86^{\rm{o}}$, $b = -36^{\rm{o}}$, then the tip of the longest simulated tail would be at
$\ell_{t1} = 125^{\rm{o}}$, $b_{t1} = -52^{\rm{o}}$.     
For comparison, the tip of the \ppa's longest tail is located at $\ell \sim 126^{\rm{o}}$, $b \sim-61^{\rm{o}}$.
Thus, the simulated structure is roughly similar to the \ppa, but is
somewhat shorter in extent 
and is approaching the midplane at a somewhat shallower angle than is the actual \ppa\ is approaching.

Next, we perform a similar analysis on the 12~Myr epoch of IVC~2.   
At this time, the brightest region of the head is
900~pc below the midplane.
Associating this location with the observed center of the \ppa's head
yields a distance between the Earth and IVC~2's head of 1530~pc.
The $y$ component of this distance is 1240~pc.
Meanwhile, the southernmost extent of the longest tail is at $z = -1500$~pc, which equates to a
latitude of $b_{t2} = -50^{\rm{o}}$.
The $x$ component of the linear span between the tip of the longest tail
and brightest region in the head is 530~pc.
Following the logic that was used on IVC~1, the longitudinal span, $\Delta \ell$,
from the tip of the longest tail to the brightest region in the head is then $25^{\rm{o}}$.
Thus, the tip of the longest tail is at $\ell_{t2} = 111^{\rm{o}}$, $b_{t2} = -50^{\rm{o}}$.
This makes IVC~2 somewhat shorter than both IVC~1 and the \ppa,
but oriented more like the \ppa\ than is IVC~1.

The angular widths of the simulated heads are also calculated.
We start with the linear width along an imaginary line that runs through the brightest part of the head and
runs perpendicular to the cloud's main axis.   This width is 190~pc for IVC~1 and 300~pc for IVC~2.
Treating this span as if it is tilted at roughly $45^{\rm{o}}$ to the midplane
yields
angular widths of 14$^{\rm{o}}$ and 11$^{\rm{o}}$, respectively.
For comparison, a similar line across
the head of the \ppa\ at its widest extent (i.e., including the southwest extension)
is approximately 10$^{\rm{o}}$ long.

\subsection{Line of Sight Velocity and Velocity Dispersion}
\label{sect:velocity}

The \ppa's first moment map is reproduced in
Figure~\ref{fig:fukui_hi_map}(b).
%
%
The main portion of the \ppa's head moves with a \los\ velocity of  $\sim-50$~km~s$^{-1}$.
Its southwest extension, where the column densities are very low,  moves at a slightly slower velocity
and its northeast side,
where the cloud follows a straight line, has both more and less extreme velocities.
The interstellar material around the head has
\los\ velocities of $\sim -10$ to $\sim 0$~km~s$^{-1}$ with respect to the LSR (see Fukui et al. 2021
Figure 5, except for the region labeled {\it{bridge}}).
Thus, the head moves at $\sim-60$ to $\sim-50$~km~s$^{-1}$ with respect to the Galactic
gas through which it is passing.
The \ppa\ has two narrow tails, one of which moves at more extreme velocities than the main part of the head,
while the other tail moves at various velocities.

For comparison,
we calculated the line of sight velocities for the 8~Myr old IVC~1 cloud and the 12~Myr year old IVC~2 cloud
from the point of view of an imagined observer
located in the Galactic midplane, 810~pc from the head of IVC~1 and 1530~pc from the head of IVC~2.
See 
Figure~\ref{fig:los_velocity}.

\begin{figure*}[htb!]		 
%

\includegraphics[trim=140 200 150 200, width=200pt, height=300pt]{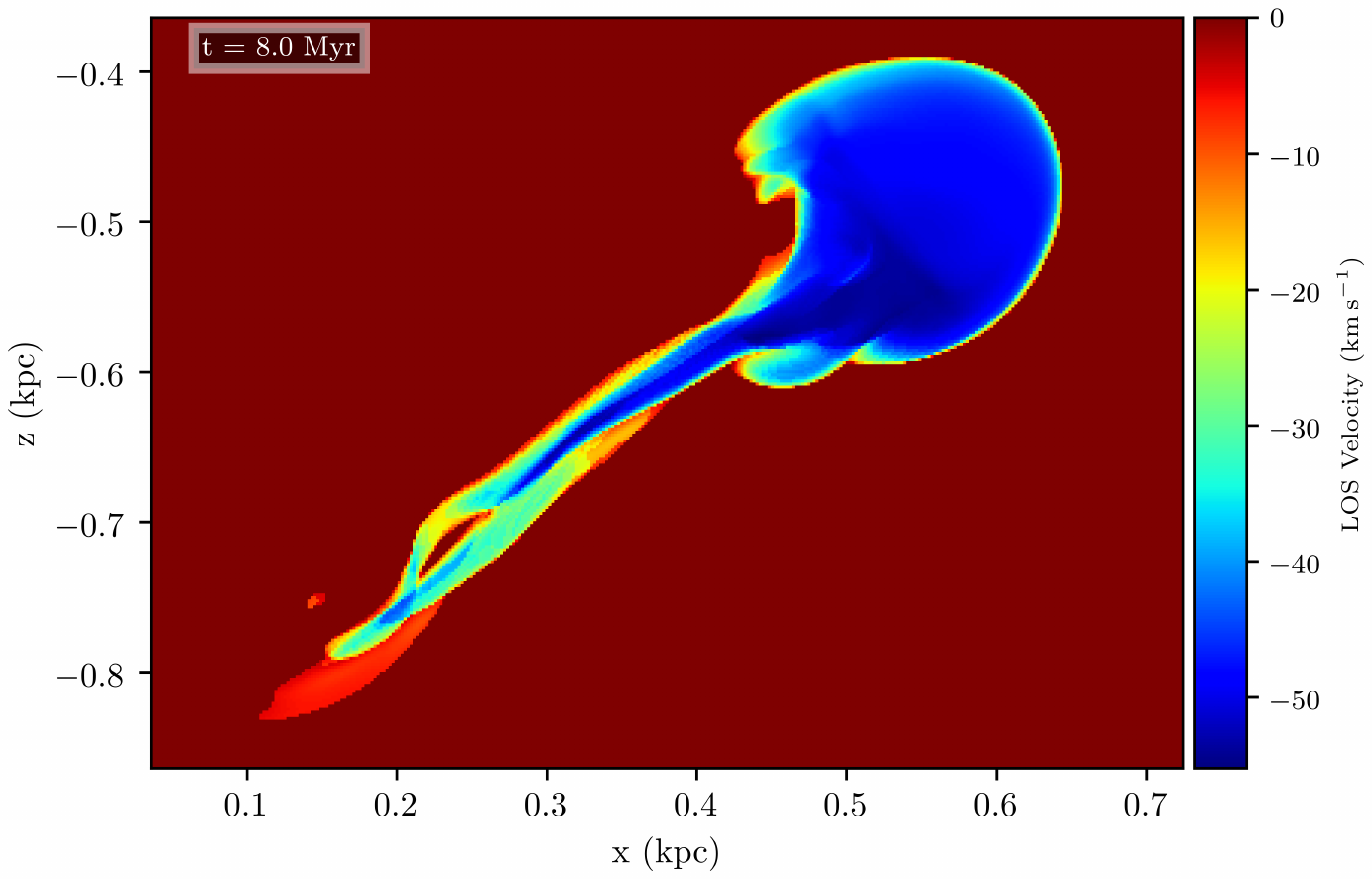}{(a)}
\includegraphics[trim=140 200 150 200, width=250pt, height=300pt]{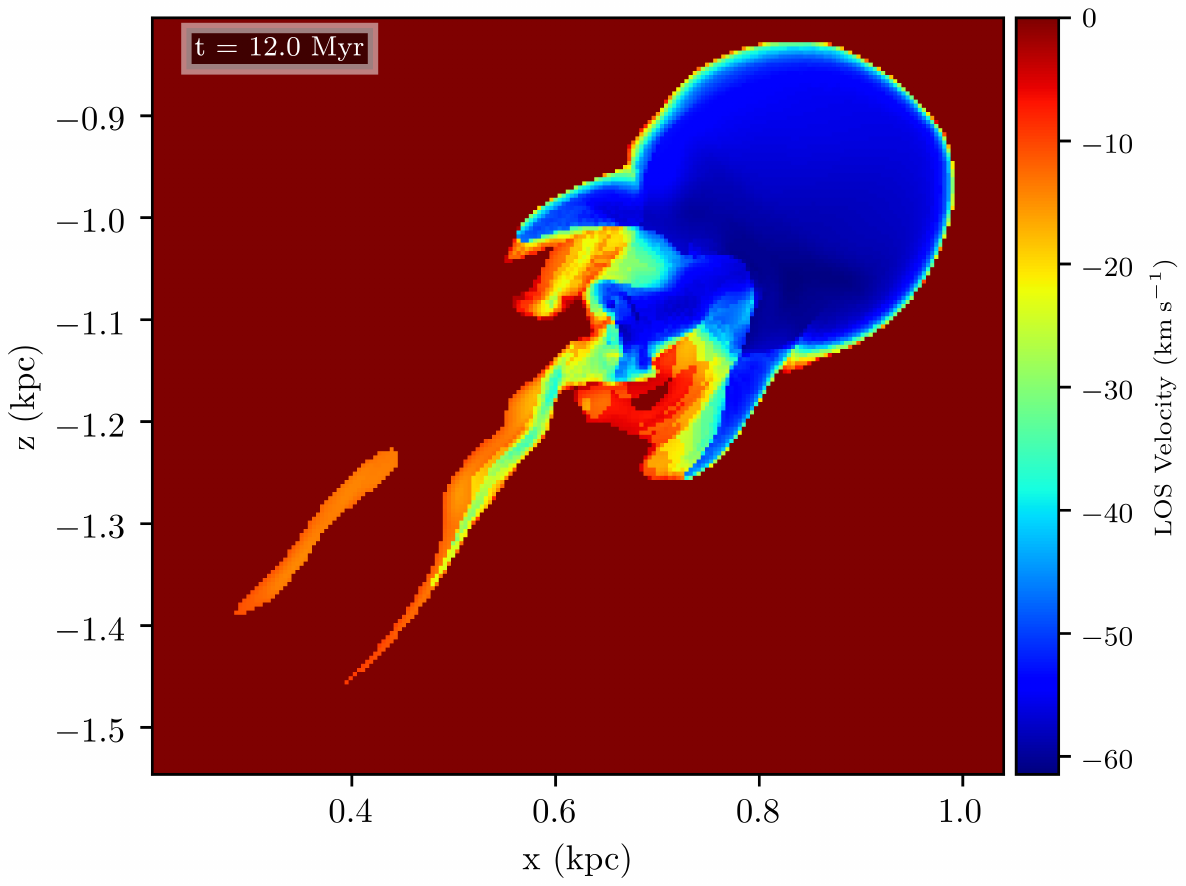}{(b)}

  \caption{Line of sight velocity maps of
    (a) IVC 1 at 8.0 Myr and
    (b) IVC 2 at 12.0 Myr.    
  }
\label{fig:los_velocity}
\end{figure*}

The velocity structure of the IVC~1 cloud exhibits similar characteristics as that of the \ppa.
The typical \los\ velocity of the main portion of IVC~1's head
is $\sim-50$~km~s$^{-1}$.
The low column density southwest margin is several km~s$^{-1}$ less extreme and
the northeast side has both more and less extreme velocity material.
The \los\ velocity of IVC~1's head is similar to that of the \ppa, but
the shape of the head and the smoothness of the velocity gradients differ from those of the \ppa.
IVC~1 has two narrow tails that overlap along the line of sight near their ends.
Like the \ppa's tails, one of IVC~1's tails approaches the viewer faster than the other
and one tail appears to be straighter than the other from the perspective of the viewer.
In both the map of the \ppa\ and the simulation,
the \los\ velocity varies along each tail, suggesting that the tails are undulating.

The head of IVC~2 travels at $\sim-57$~km~s$^{-1}$ along the line of sight, but
both the southwest and northeast margins travel at less extreme velocities.
A simple translational velocity shift of $\sim7$ to $\sim17$ ~km~s$^{-1}$ would shift the simulated head's
line of sight velocities to approximately those of the \ppa's head.
Like the other model, IVC~2 differs from the \ppa\ in both the shape of the simulated head and
the smoothness of the velocity gradients.   The lack of faster material on the northeast side of IVC 2's head is an additional
difference with the \ppa.
IVC~2's trailing gas shows a clear velocity gradient from northwest to southeast.
No part of the trailing gas moves toward the viewer faster than the center of the head does,
in contrast
with one tail of the \ppa\ and in contrast with one tail of IVC~1.
Of the two simulations, IVC~1 is more similar to the \ppa\ in regards to velocity structure.

We have created a dispersion map for the \ppa\ from the GALFA-\hone\ data (Peek et al. 2011) analyzed in Fukui et al. (2021).
For comparison with it, we calculated the line of sight velocities of the cooler material in the
IVC~1 and IVC~2 simulations.
Since the clouds are cooler than the background gas, this selection criterion
cuts out the background gas.
      Figure~\ref{fig:simulated_dispersion} presents the
      velocity dispersion maps for the heads of the \ppa\ and the two simulated clouds.
      Each map has a similar range of velocity dispersions:
2 to 10 km s$^{-1}$ for the head of the \ppa,
compared with 3 to $\sim12$ km s$^{-1}$, with an envelope of higher dispersion gas for the head of IVC~1
and 4 to $\sim12$ km s$^{-1}$, with an envelope of higher dispersion gas for the head of IVC 2.
The dispersion in the middle of the \ppa's head is around 5~km~s$^{-1}$ which is similar to
the median dispersions in IVC~1 are and IVC~2 (i.e., $\sim6$~km~s$^{-1}$).
The regions of greatest dispersion in the \ppa's head are scattered spots and the northeastern ridge.
The head's dispersion generally decreases from the (faster-moving) northeast side to the (slower-moving)
southwestern extension.
In comparison, the simulated cloud heads are encircled with high dispersion gas
and the dispersion decreases smoothly
to lower values in the center.

\begin{figure*}[htb!]		 
%

\includegraphics[trim=140 200 150 200, width=240pt, height=300pt]{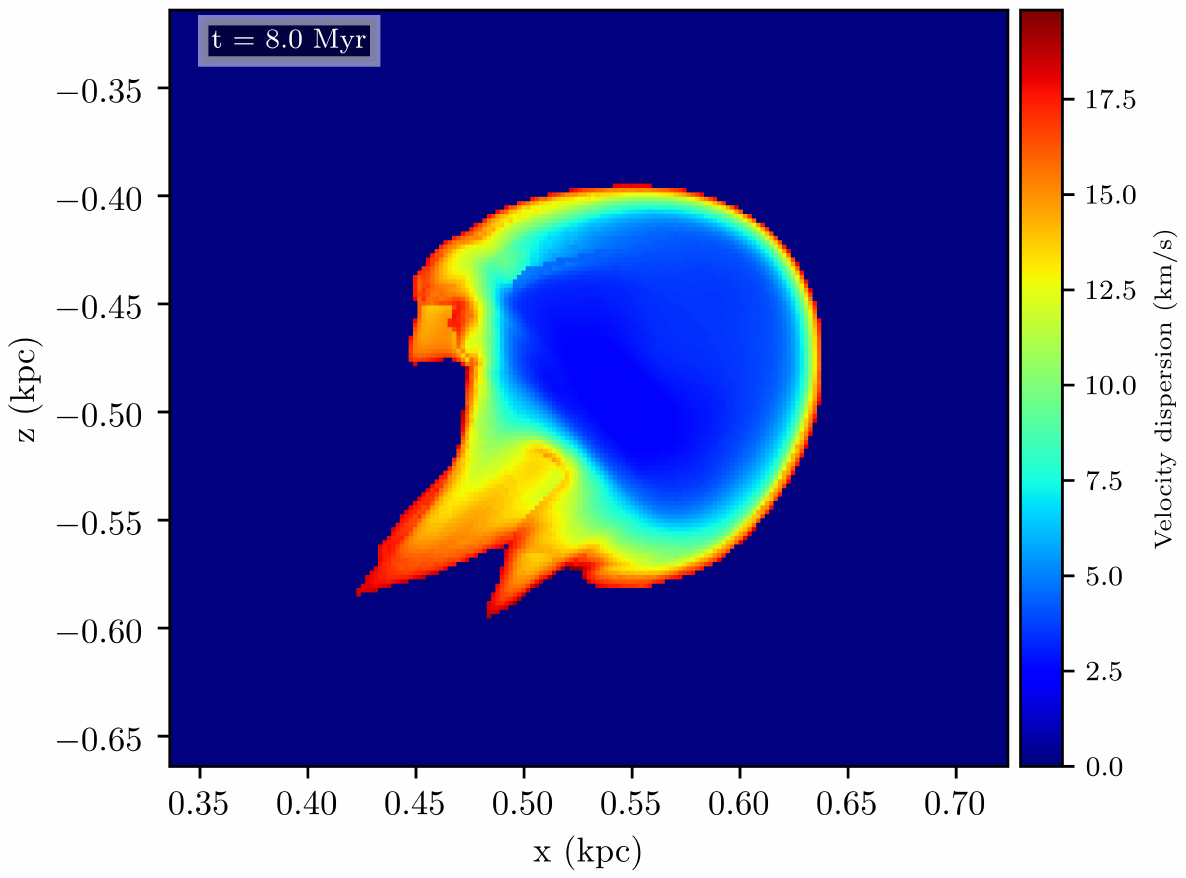}{(a)}
\includegraphics[trim=140 200 150 200, width=240pt, height=300pt]{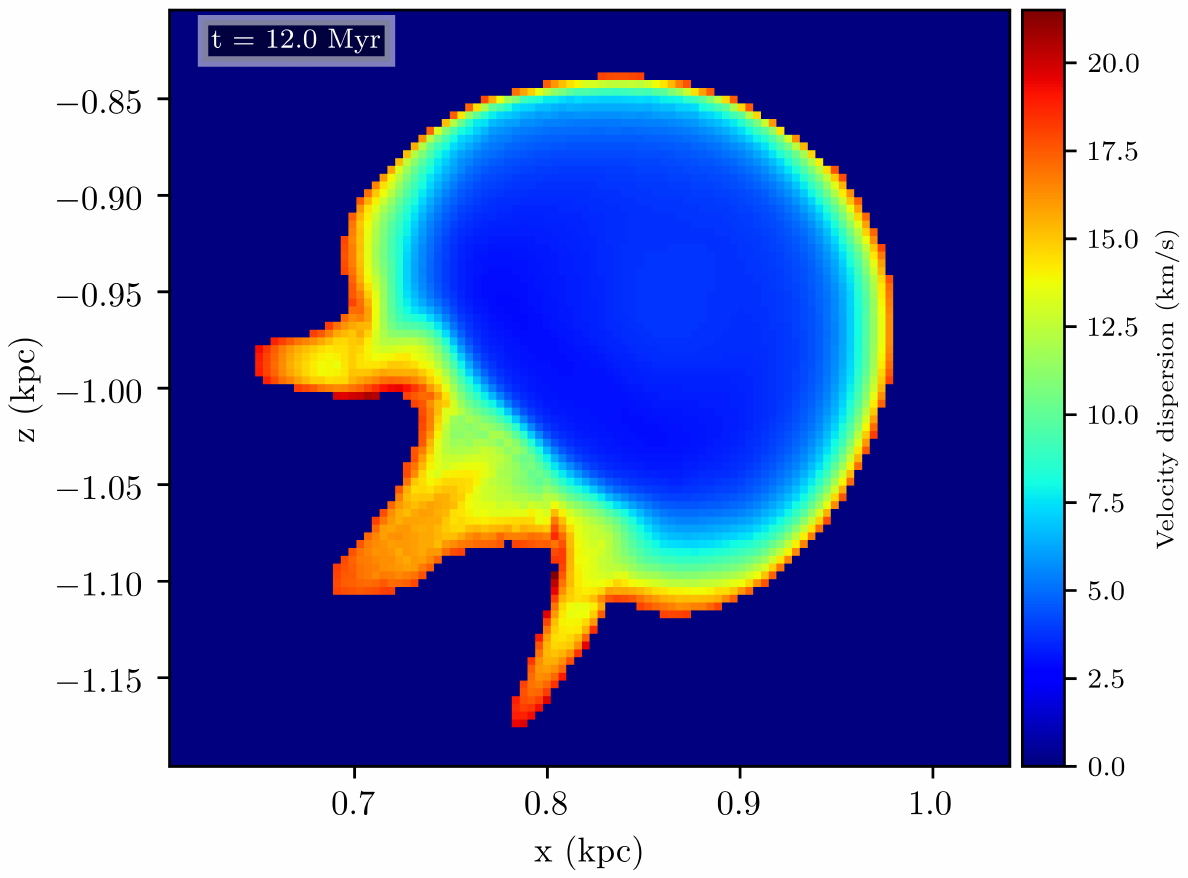}{(b)}

  \caption{Velocity dispersion maps of the heads of (a) IVC 1 at 8.0 Myr and (b) IVC 2 at 12.0 Myr.
  }
\label{fig:simulated_dispersion}
\end{figure*}

\subsection{Predicting the Future for the \ppa}

Figures~\ref{fig:ivc1_density_slices}(d)-(f)
and \ref{fig:ivc2_density_slices}(e) - (i)
portray the future evolution of the simulated clouds, showing that they collide with the Galactic disk and dissipate.
%
The IVC~1 cloud disappears entirely by 20~Myr and the IVC~2 cloud disappears entirely by 21~Myr.
The clouds do not reach $|z| < 100$~pc, but their interactions with the ISM
compress both Galactic gas and formerly cloud gas into a thin
layer of upward moving gas that does.
In the IVC~1 simulation, this compressed layer is able to dislocate the midplane gas.
In essence, the clouds transfer their material, momenta, and kinetic energy to the disk.
In the case of IVC~1, there is $\sim7 \times 10^{50}$~erg of transferred kinetic energy and
in IVC~2, there is $\sim3 \times 10^{51}$~erg of transferred kinetic energy, i.e., roughly
one to several supernova(e) worth of energy.   

\section{Discussion}
\label{sect:discussion}

The long, narrow, and relatively smooth tails of the \ppa\ are fairly unique among fast-moving clouds, 
raising the question of why these characteristics developed on the \ppa\ but not other clouds.
It may be the case that the cloud's Mach number and Reynolds number
are important.
Although the \ppa's line of sight speed is around $-50$~km~s$^{-1}$,
its total speed is estimated at $\sim100$~km~s$^{-1}$.
The cloud is traveling through the Reynold's Layer, whose temperature is around $10^4$~K
(Reynolds 1990).  
At this temperature, the sound speed is around 10 km~s$^{-1}$,
which is a small fraction of the cloud's total speed.
Therefore, the \ppa\ should be highly supersonic and should instigate a bow shock.
The effect of a bow shock front is to accelerate the material behind the bow shock.
This reduces the relative speed between the cloud and the material immediately around it.
A bow shock forms in each of the simulations, as well.
As expected, in each simulation, the bow shock decreases the velocity contrast between the cloud and the gas immediately around it.
For example, the gas around IVC~1's head at 5~Myr has been sped up so much that the velocity contrast between
the head and it is only 40~km~s$^{-1}$.
The cloud's tails are also in the accelerated but calm region far behind the bow shock.
%

The conditions in the Reynold's Layer are in contrast with the halo and circumgalactic gas surrounding many HVCs.
The halo and circumgalactic medium have temperatures of
$\sim2 \times 10^6$~K (Henley \& Shelton 2015; Nakashima et al. 2018)
and so the sound speed is approximately
$\sim14$ times larger than that of the $10^4$~K Reynold's layer.
Hotter components have also been found (Das et al. 2019), for which the sound speed is even higher.
Even typical HVCs with speeds of $\gtrsim 100$~km~s$^{-1}$ are subsonic or only marginally supersonic in $T = 2 \times 10^6$~K gas.
Only the fastest HVCs would create bow shocks in this gas.
%
%

The relative velocity, temperature, and density affect the Reynolds number, $Re$, which theoretically
governs whether a moving object's wake is turbulent or laminar.
From Benjamin (1999),
$Re = LV/\nu_{\rm{eff}}$, where $L$ is the length of the object,
$V$ is the velocity (for which we use the relative velocity), and $\nu_{\rm{eff}}$ is the effective viscosity,
%
%
which equates to $6 \times 10^{19} \, (T/10^4\,{\rm{K}})^{5/2} (0.01\,{\rm{cm}}^{-3}/n)$ in the absence of magnetic fields,
where $T$ is the temperature and $n$ is the density.
We evaluate $Re$ for the head and tail for each cloud.
Our calculations of $Re$ for the heads use earlier epochs because the conditions around each head at earlier times
set the stage for turbulence downstream at later times and
because using earlier epochs enables us to avoid a small density wave that travels through each domain.
We use 5~Myr for IVC~1 and 11~Myr for IVC~2.
At 5~Myr, $L$
the width of the head 
is 210~pc,
the velocity contrast between the head and the
surrounding gas is $V = 40$~km~s$^{-1}$,
the average $T$ in the material surrounding the head is
$5.0 \times 10^5$~K,
and the density of atoms and atomic nuclei is $2.37 \times 10^{-4}$~cm$^{-3}$.
This gas is hotter and more rarified than the Reynold's Layer,
because
simulating hydrostatic balance in the thick disk due to thermal pressure and constrained by a realistic midplane density
requires somewhat higher temperatures and lower densities than those of the Reynold's Layer.
From these values, $\nu_{\rm{eff}} = 4.5 \times 10^{25}$ and $Re = 58$ for the head of IVC~1.
We perform similar calculations for 
the head of IVC~2 at 11~Myr.   At this time,
$\nu_{\rm{eff}} = 1.5 \times 10^{26}$ and $Re = 34$.
The values of $\nu_{\rm{eff}}$ and $Re$ are evaluated for the tail at the fiducial epochs, yielding
$\nu_{\rm{eff}} = 7.6 \times 10^{25}$ and $Re = 4.8$ for IVC~1 
and $\nu_{\rm{eff}} = 1.1 \times 10^{26}$ and $Re = 5.4$ for IVC~2.
All of the $Re$ values are very low, which portends laminar flow rather than turbulent flow.
Even if the value of $n_{-2}  \, T_4^{-5/2}$ in the medium around the \ppa\ were a couple of orders larger
than in these simulations, the Reynolds numbers would remain lower than those of
turbulent flow.

These simulations do not model the magnetic field.   On one hand, Benjamin (1999) indicates that magnetic fields
decrease the effective viscosity substantially, thus increasing the Reynolds number substantially.    On the other hand,
simulated IVCs and HVCs that include magnetic fields tend to develop obvious tails, too.   See Santillan et al. (1999),
Kwak et al. (2009), Jel\'{i}nek \& Hensler (2011), Kwak et al. (2011), and Galyard \& Shelton (2016) for examples.

The viewing geometry may also play a role in explaining the difference between the \ppa's morphology and those of
other fast-moving clouds.
The tails of the \ppa\ are obvious because they are oriented approximately perpendicular to the line of sight.
If, in contrast, the \ppa\ were to be moving directly toward the viewer, its head would overlap its tail.  The cloud
would look like a blob from that point of view.  
If the cloud were to be observed at an intermediate viewing angle, the tails would appear foreshortened, with
velocity gradients from the head to the end of the tail.   This is more similar to the head-tail
HVCs (Br\"{u}ns et al. 2000).

    Regarding the cloud's direction of motion, from the sweep of the cloud and its relative proximity to the Sun,
    it appears that the cloud is moving from the outer Galaxy to the inner Galaxy.   It would have passed below
    the Perseus Arm and be in the process of passing by the Orion-Cygnus (or Local) Arm.   Its estimated speed in
    the direction of Galactic rotation is slower than that of Milky Way material.
    Therefore, the thick disk ISM is broadsiding the cloud and should be accelerating the cloud in the direction
    of Galactic rotation.    A current topic of interest in studies of HVCs asks
    whether the angular momentum vectors of infalling clouds are somewhat aligned with that of the Milky Way.
    The \ppa\ may be an interesting case for further examination with regards to this question.

\section{Summary}
\label{sect:summary}

We present simulations of the \ppa, an IVC with unusually long twin tails that is thought to have extragalactic origins.
Our simulations track the past, present, and future evolution of the cloud.
Each simulation begins with a spherical cloud located
$\sim1$ to 2~kpc
from the midplane and
moving obliquely toward the Galactic disk.
As the simulated clouds move toward the disk, 
they develop long,
bifurcated tails.
Each simulated head remains intact until it
gets within $\sim 150$~pc of the midplane,
whereupon it is crushed by its collision with the Galactic disk.   
The simulated IVCs dissipate and are absorbed by the Galaxy.

The current distance between the Earth and the \ppa\ is not well known.  Nor are the cloud's initial location, mass, or velocity.
Therefore, we developed simulational models of the cloud located at a nearer distance and a farther distance.
The observed velocity and \hone\ intensity of the head match the simulated values in both of these models.
%
%
These are Simulations IVC~1 and IVC~2.
During the epoch when the IVC~1 cloud looks most similar to the \ppa\ (i.e., 8~Myrs after the beginning of the simulation),
the head of the cloud is 810~pc from Earth
and during the epoch when the IVC~2 cloud looks most similar to the \ppa\ (i.e., 12~Myrs after the beginning of the simulation),
the head is 1530~pc from Earth.
Both of these
distances are within the known constraints on the distance to the head of the \ppa.
%
Note that the quoted epoch age of either model merely corresponds to the length of simulated time
since the simulation began.
The actual cloud would be older than such an epoch age,
because some amount of real time must have elapsed while the cloud was forming 
and before it reached the location modeled at the beginning of the simulation.

Eight megayears into the IVC~1 simulation,
the cloud's tails are approaching the angular length of the \ppa.
One tail
appears straight on the plane of the sky from the observer's point of view.  Its \los\ velocity
varies along its length, which indicates that the tail is wavering.
The other tail has a bend in it.   In addition, its line-of-sight velocity also varies along the length of the tail, indicating
that the tail is wavering.
The \ppa, similarly has one curvy tail and one straight tail.
As in the simulation, the \los\ velocity varies along the length of the straight tail 
in the map of the \ppa, suggesting 
that it is wavering.

The tails of IVC~1 have greater angular lengths
than those of IVC~2.
In this regard, IVC~1 provides better morphological similarity to the \ppa.    
However, when it comes to the head of the cloud, IVC~2 is the more appropriate model because its head becomes flatter.
IVC~2's head is flattest
after the time when its tails most resemble those of the \ppa.
The flattening is caused by the cloud's collision with Galactic disk gas.
For comparison,
the head of the \ppa\ is flat, but the effect is greater and  
is farther to the east than the flattening of IVC~2.
Based on the trends seen in these simulations, it is likely that 
greater and more eastern flattening could have come about if,
as Fukui et al. (2021) suggested,
the head of 
the \ppa\ had collided with a slightly dense, high altitude interstellar gas cloud.



Smooth, extended tails are not common on other IVCs or HVCs in our Galaxy, but appear on the \ppa, these simulations, and
preliminary simulations performed for this project.   An explanation for the smoothness of the tails is that the
Reynolds number is low (see Section~\ref{sect:discussion}), foretelling a nonturbulent flow.   
Another point to consider is that if the \ppa\ had been located directly above the solar neighborhood rather than at
$b = -36^{\rm{o}}$ to $-61^{\rm{o}}$, then geometrical foreshortening would have caused it to look like globular, more like
other IVCs and HVCs.



The simulated clouds do not survive their inevitable impacts with the Galactic disk.
It is reasonable to think that the \ppa\ will have the same fate.
%
In that case, its 
mass, momentum, and kinetic energy will be given over to the Galactic disk and the gas just above it.
The transferred kinetic energy could be equivalent to that of $\sim1$ to $\sim6$ supernova(e) explosion(s).

\section*{Acknowledgements}
We appreciate the suggestions offered by the anonymous referee.   They have made this a better paper.
We acknowledge and appreciate
Dr. Shan-Ho Tsai for her assistance with the computer clusters.
These simulations were performed on computers at the Georgia Advanced Computing Resource Center at
the University of Georgia.     
The FLASH code used in this work was in part developed by the DOE-supported ASC/Alliance Center
for Astrophysical Thermonuclear Flashes at the University of Chicago. 
The simulational work was supported through grant NNX13AJ0G through the NASA ATP program.
We acknowledge Takahiro Hayakawa for his contributions to the preparation of the observational figures.
The observational work was financially supported by Grants-in-Aid for Scientific Research (KAKENHI) of the
Japan Society for the Promotion of Science (JSPS) through grant numbers 15H05694, 20H01945, and 21H00040.

\pagebreak


\begin{thebibliography}{0}



\bibitem[Armillotta et al.(2017)]{armillotta_etal_2017}
  Armillotta, L., Fraternali, F., Werk, J., K., Prochaska, J. X., \& Marinacci, F.\ 2017, \mnras, 470, 114
  

\bibitem[Benjamin(1999)]{benjamin_1999}Benjamin, R. 1999, ASPC, 166, 147



\bibitem[Br{\"u}ns et al.(2000)]{bruns_etal_2000}
  Br{\"u}ns, C., Kerp, J., Kalberla, P.~M.~W., \& Mebold, U.\ 2000, \aap, 357, 120


\bibitem[Br{\"u}ns et al.(2001)]{bruns_etal_2001} Br{\"u}ns, C., Kerp, J., \& Pagels, A. 2001, A \& A, 370, L26
  
  

\bibitem[Centurion et al. (1994)]{centurion_etal_1994}
  Centurion, M., Vladilo, G., de Boer, K. S., Herbstmeier, U., \& Schwarz, U. J. 1994, A $\&$ A, 292, 261



  
  \bibitem[Danly (1989)]{danly_1989} Danly, L. 1989, \apj, 342, 785

  \bibitem[Das etal.(2019)]{das_etal_2019} Das, S., Mathur, S., Gupta, A., Nicastro, F., \& Krongold, Y.
  2019, ApJ, 887, 257
  
  \bibitem[de Avillez \& Breitschwerdt(2007)]{deavillez_breitschwerdt_2007}
  de Avillez, M. A. \& Breitschwerdt, D. 2007, \apj, 665, L35
  


  \bibitem[Fitzpatrick \& Spitzer(1997)]{fitzpatrick_spitzer_1997} Fitzpatrick, E. L. \& Spitzer, L. 1997, ApJ, 475, 623
  
  \bibitem[For et al.(2014)]{for_etal_2014} For, B.-Q., Staveley-Smith, L., Matthews, D., $\&$ McClure-Griffiths, N. M. 2014,
    ApJ, 792, 43

   \bibitem[Fox et al.(2013)]{fox_etal_2013} Fox, A. J., Richter, P., Wakker, B. P., et al. 2013, the Messenger, 153, 28

  \bibitem[Fox et al.(2014)]{fox_etal_2014} Fox, A. J., Wakker, B. P., Barger, K. A., et al. 2014, ApJ, 787, 147

  \bibitem[Fryxell, B., et al. (2000)]{fryxell_2000}
  Fryxell, B., et al.\ 2000, \apjs, 131, 273

\bibitem[Fukui et al.(2015)]{fukui_etal_2015} Fukui, Y., Torii, K., Onishi, T., et al.\ 2015, \apj, 798, 6



  \bibitem[Fukui et al.(2021)]{fukui_etal_2021} Fukui, Y., Koga, M., Maruyama, S., et al. 2021, PASJ, 73, S117
  
  \bibitem[Galyardt \& Shelton(2016)]{galyardt_shelton_2016} Galyardt, J. \& Shelton, R.~L.\ 2016, \apjl, 816, L18

  \bibitem[Gritton et al.(2014)]{gritton_etal_2014} Gritton, J.~A., Shelton, R.~L., \& Kwak, K.\ 2014, \apj, 795, 99

  \bibitem[Haffner et al.(2001)]{haffner_etal_2001} Haffner, L. M., Reynolds, R. J., \& Tufte, S. L. 2001, ApJ, 556, L33

  \bibitem[(Hartmann \& Burton(1997)]{hartmann_burton_1997} Hartmann, D. \& Burton, W. B. 1997,
  Atlas of Galactic Neutral Hydrogen (Cambridge: Cambridge Univ. Press)
  
  \bibitem[Heitsch \& Putman(2009)]{heitsch_putman_2009} Heitsch, F. \& Putman, M.~E.\ 2009, \apj, 698, 1485

  \bibitem[Henley \& Shelton(2015)]{henley_shelton_2015} Henley, D. B. \& Shelton, R. L. 2015, \apj, 808, 22

  \bibitem[Hernandez et al.(2013)]{hernandez_etal_2013} Hernandez, A.~K., Wakker, B.~P., Benjamin, R.~A., et al. 2013, \apj, 777, 19

  \bibitem[Jel\'{i}nek \& Hensler(2011)]{jelinek_hensler_2011} Jel\'{i}nek, P. \& Hensler, G. 2011, CoPhC, 182, 1784


  \bibitem[Kalberla et al.(2005)]{kalberla_etal_2005} Kalberla, P. M. W., Burton, W. B., Hartmann, D., et al.
  2005, A \& A, 440, 775
  
  \bibitem[Kwak, et al.(2009)]{kwak_etal_2009} Kwak, K., Shelton, R. L., Raley, E. A.  2009,  ApJ, 699, 1775

  \bibitem[Kwak et al.(2011)]{kwak_etal_2011} Kwak, K., Henley, D. B., \& Shelton, R. L.\ 2011, ApJ, 739, 30

  \bibitem[Kuntz \& Danly(1996)]{1996ApJ...457..703K} Kuntz, K.~D. \& Danly, L.\ 1996, \apj, 457, 703 



  \bibitem[MacNeice et al.(2000)]{macneice_etal_2000}
  MacNeice, P., Olson, K. M., Mobarry C.,
  de Fainchtein R., Packer C.\ 2000, CoPhC, 126, 330

  \bibitem[Magnani, Blitz, \& Mundy (1985)]{magnani_etal_1985} Magnani, L., Blitz, L., \& Mundy, L. 1985, ApJ, 295, 402
  

\bibitem[Marasco et al.(2019)]{marasco_etal_2019} Marasco, A., et al. 2019, A \& A, 631, A50
    

  \bibitem[Moehler et al. (1990)]{moehler_etal_1990} Moehler, S., Heber, U., and de Boer, K. S. 1990, A$\&$A, 239, 265
  
  \bibitem[Nakashima et al.(2018)]{nakashima_etal_2018}Nakashima, S., Inoue, Y., Yamasaki, N., et al. 2018, ApJ, 862, 34

  \bibitem[Nidever et al.(2008)]{nidever_etal_2008} Nidever, D. L., Majewski, S. R., \& Burton, W. B. 2008, ApJ, 679, 432

  \bibitem[Orlando, et al.(2003)]{orlando_etal_2003} Orlando, S., Peres, G., Preal, F., et al. 2003, MSAIS 4, 82

  \bibitem[Parker(2019)]{parker_2019} Parker, M. C. 2019, ``Intermediate-Velocity Clouds Approaching the Galactic Plane''
    Masters Thesis, Department of Physics and Astronomy, University of Georgia, Athens, GA USA


  \bibitem[Peek et al.(2011)]{peek_etal_2011} Peek, J.~E.~G., et al. 2011, ApJS, 194, 20

  \bibitem[Pl\"{o}ckinger \& Hensler(2012)]{plockinger_hensler_2012} Pl\"{o}ckinger, S., \& Hensler, G.\ 2012,
  \aap, 547, A43

  \bibitem[Putman et al.(2011)]{putman_etal_2011} Putman, M. E., Saul, D. R., $\&$ Mets, E. 2011, MNRAS, 418, 1575
 
  \bibitem[Reynolds(1990)]{reynolds_1990} Reynolds, R. J. 1990, ApJL, 349, L17 


  \bibitem[Richter et al.(2013)]{richter_etal_2013}Richter, P., Fox, A. J., Wakker, B.~P.,  et al. 2013, ApJ, 772, 111
  
  \bibitem[Richter(2017)]{richter_2017}Richter, P. 2017, ASSL, 430, 15
  
  \bibitem[R{\"o}hser et al.(2016)]{rohser_etal_2016} R{\"o}hser, T., Kerp, J., Lenz, D., \& Winkel, B.\ 2016, \aap, 596, A94

  \bibitem[Sander \& Hensler(2019)]{sander_hensler_2019} Sander, B. \& Hensler, G. 2019, MNRAS, 490, L52

  \bibitem[Sander \& Hensler(2020)]{sander_hensler_2020} Sander, B. \& Hensler, G. 2020, MNRAS, 501, 5330
  
  \bibitem[Santill\'{a}n et al.(1999)]{santillan_etal_1999}Santill\'{a}n, A., Franco, J., Martos, M., \& Kim, J. 1999, ApJ, 515, 657

  \bibitem[Santill\'{a}n et al.(2004)]{santillan_etal_1999}Santill\'{a}n, A., Franco, J., \& Kim, J. 2004, JKAS, 37, 233

  \bibitem[Smoker et al.(2011)]{smoker_etal_2011}Smoker, J. V., Fox, A. J., $\&$ Keenan F. P. 2011, MNRAS, 415, 1105

  \bibitem[Takahira et al.(2014)]{takahira_etal_2014}Takahira, K., Tasker, E. J., \& Habe, A. 2014, ApJ, 792, 63
    
  \bibitem[Tepper-Garc\'{i}a \& Bland-Hawthorn(2018)]{tepper_garcia_bland_hawthorn_2018} Tepper-Garc\'{i}a, T. \& Bland-Hawthorn, J. 2018 MNRAS, 473, 5514
  
  \bibitem[Torii et al.(2017)]{torii_etal_2017} Torii, K. et al. 2017, ApJ, 835, 142

 
  \bibitem[Wakker(2001)]{2001ApJS..136..463W} Wakker, B.~P.\ 2001, \apjs, 136, 463

  \bibitem[Wakker \& van Woerden(1997)]{wakker_vanwoerden_1997} Wakker, B.~P. \& van Woerden, H.\ 1997, \araa, 35, 217


  \bibitem[Westmeier(2018)]{westmeier_2018} Westmeier, T. 2018, MNRAS, 474, 289

  \bibitem[Yamamoto et al.(2003)]{yamamoto_etal_2003}Yamamoto, H., Onishi, T., Mizuno, A., \& Fukui, Y. 2003, ApJ, 592, 217
  
  
%
	

\end{thebibliography}
\end{document}